\documentclass[12pt]{article}
\usepackage{feynmp}
\usepackage{slashed}
\usepackage{graphicx}
\usepackage{epstopdf}
\usepackage{subfigure}
\usepackage{amsmath}
\usepackage{amsfonts}
\usepackage{amssymb}
\usepackage{color}
\usepackage{mathrsfs}
\usepackage[colorlinks=true, citecolor=blue, urlcolor=blue]{hyperref}
\hypersetup{linkcolor=blue}

\newcommand{\beq}{\begin{equation}}
\newcommand{\eeq}{\end{equation}}
\newcommand{\bea}{\begin{eqnarray}}
\newcommand{\ea}{\end{eqnarray}}
\newcommand{\barr}{\begin{array}}
\newcommand{\earr}{\end{array}}

\def\bea{\begin{eqnarray}}
\def\eea{\end{eqnarray}}

\def\nn{\nonumber\\}

\newcommand{\lb}{{\langle}}
\newcommand{\rb}{{\rangle}}

\newcommand{\calo}{{\cal O}}

\begin{document}

\begin{titlepage}

\setcounter{page}{1} \baselineskip=15.5pt \thispagestyle{empty}

\vfil

\begin{center}

\def\thefootnote{\fnsymbol{footnote}}

\begin{changemargin}{0.05cm}{0.05cm} 
\begin{center}
{\huge \bf Non-Gaussianities in Dissipative }
\end{center} 
\begin{center}
{\huge \bf EFT of Inflation Coupled to a Fluid}
\end{center}
\end{changemargin}

~\\[0.5cm]
{\large Gustavo J. Turiaci$^1$ and Matias Zaldarriaga$^2$}
\\[0.5cm]

{\normalsize { \sl $^{1}$ Physics Department, Princeton University, Princeton, NJ 08544, USA.}}\\
\vspace{.3cm}

{\normalsize { \sl $^{2}$Institute for Advanced Study, Einstein Drive,  Princeton, NJ 08540, USA.}}\\
\vspace{.3cm}

\end{center}

\vspace{1.8cm}

 \vspace{0.2cm}
\begin{changemargin}{01cm}{1cm} 
{\small  \noindent 
\begin{center} 
\textbf{Abstract}
\end{center} 
\noindent We studied models of inflation with a preferred clock specifying the end of inflation and giving the curvature perturbations, coupled with another non-equivalent clock that at late times defines the same frame and do not contribute to the density perturbations. This can happen in the framework of dissipative EFT of inflation where the additional degrees of freedom include a fluid developing sound waves propagating with sound speed $c_{sr}$. The fluid defines a preferred frame comoving with it.  The paradigmatic example of this is the warm inflation scenario. We studied the dynamics of this systems during inflation and the three-point function. We saw that in the strong dissipation regime the nonlinear parameter induced by the new terms is $|f_{\rm NL}| \sim 1/c_{sr}^2$, not enhanced by the dissipation parameter which enters the two-point function. We checked that the squeezed limit of the three-point function still satisfies the consistency condition with corrections of order $\calo(k_L^2/k_S^2)$. We computed the Planck constraints for the case of warm inflation obtaining a bound of $\gamma\lesssim 10^5 H$ for the clock coupled to radiation. For decreasing sound speed the bound decreases. We also checked that the shape of the three-point function corresponding to the model studied here is a mixture between equilateral and orthogonal with a small local component, which is more consistent with Planck's results.}
\end{changemargin}
 \vspace{0.3cm}
\vfil
\begin{flushleft}
\today
\end{flushleft}

\end{titlepage}


\section{Introduction}

~~~~~Recently, in \cite{diseft} the effective field theory (EFT) of single-field inflation was generalized to include dissipative effects. In these kind of models one thinks of a preferred clock breaking time diffeomorphisms, as in the EFT of single-field inflation, coupled to a dissipative sector composed of additional degrees of freedom (ADOF). The EFT describing these kind of systems is different from the one describing multi-field inflation \cite{multieft}. In multi-field inflation the extra degrees of freedom are directly affecting the dynamics at late times after the mode exits the horizon. They can modify the reheating time and therefore the duration of inflation for example. In the dissipative EFT of inflation the ADOF only couple to the clock at early times before the mode we observe exits the horizon. Although they share the presence of an extra sector coupled to the clock the features of these two models are very different \cite{diseft,multieft}.

In this paper we will still concentrate on the case of dissipative effects of inflation with a preferred clock but we will consider systems where the ADOF develop a time-like four vector $\lb u^\mu_\calo\rb\neq 0$ with non-vanishing expectation value, thus defining a second preferred frame. To stay within the framework of dissipative EFT of inflation this frame must coincide with the preferred frame fixed by the first clock at late times. We will see that the presence of such a four-vector's dynamics produces an interaction which is non-local in time on a scale of a Hubble time \footnote{We will call the interaction non-local because it involves perturbations at previous times as opposed to the interaction studied in \cite{diseft} which is ``local" in time. This should not be confused with the notion of local shape of the three-point function.}. Therefore some of the approximations used in \cite{diseft} are no longer valid. The phenomena driving this non-local behavior are sound waves in the fluid parameterized by a speed of sound $c_{sr}$. 

An example of this kind is the warm inflation scenario \cite{warm}. The inflaton is a scalar field coupled to a fluid (radiation) thus defining two preferred frames, the one comoving with the fluid and the one where the inflaton field is homogeneous. In this model the radiation is in a thermal state and the decay of the inflaton into fluid particles results in a strong dissipative regime for the inflaton. Interactions between the fluid and the field can generate non-Gaussian correlations \cite{moss,moss2}.

In these cases one is between the two paradigms. There are more than one preferred frame on the one hand but on the other hand they coincide at late times. To decide if these models resemble more one paradigm or the other (multi-field EFT or dissipative EFT) one can look at the most prominent features of both. For example, multi-field models can  present a non-trivial squeezed limit for the three-point function. For single-field or dissipative EFT there is a consistency condition constraining the parameter 
\beq
f^{\rm sq}_{\rm NL} \equiv \lim_{k_3\rightarrow0} \frac{5}{6} \frac{F(k_1,k_2,k_3)}{P_\zeta(k_1)P_\zeta(k_2)+P_\zeta(k_1)P_\zeta(k_3)+P_\zeta(k_3)P_\zeta(k_2)},
\eeq
where $F(k_1,k_2,k_3)$ is proportional to the three-point function of curvature perturbations $\zeta$ and $P_\zeta$ its power spectrum, to be 
\beq
f_{\rm NL}^{\rm sq} = \frac{5}{12}(n_s-1),
\eeq
where $n_s$ is the tilt of the spectrum measuring deviations form scale-invariance. This quantity in these models is of the order of the slow-roll parameters. This was proven in \cite{maldacena, consist1,consist2, diseftcon}. The argument is quite general and relies on the fact that long-wavelength modes can be locally gauged away and therefore do not interact with short-wavelength modes. This is true whenever there is a preferred clock driving inflation, whether single-field or dissipative. 

Nevertheless, in the warm inflation scenario it was found that this condition is violated and large local non-gaussianities can develop \cite{moss}, signaling a departure from the dissipative EFT paradigm. Here we will show that the computation of the three-point function in these models actually gives a result resembling more closely the one of \cite{diseft} with a trivial squeezed limit satisfying the consistency condition of \cite{diseftcon}. The reason will become clear below and it is due to the fact that when a mode exits the horizon fluid perturbations vanish since they are sourced by gradients. Furthermore, the ADM parameters $N,N^i$ go to their unperturbed values allowing a long-wavelength mode to be eliminated by a change of coordinates. The case studied here is richer though because different regimes appear depending on whether the long mode has exited either the Hubble or sound horizon when the short mode freezes. 

This issue is not trivial since recent analysis of Planck data \cite{planckinf,planckng} used this quasi-local shape to constrain the warm inflation model. We will revise the constraints imposed by the data using the calculations presented in this paper. We will find a bound on the dissipation coefficient of $\gamma\lesssim10^5 H$ with a shape mostly equilateral/orthogonal and with a small local contribution. This shape is in better agreement with Planck measurements than previous analysis suggested.  

 In section \ref{sec:eft} we review the basics of the EFT treatment of dissipative inflation with a preferred clock. We give a summary of the ingredients of warm inflation relevant for this work. We also give a practical way of obtaining the most general interaction for the most general possible model of inflation with a preferred clock coupled to ADOF with a preferred frame. In section \ref{sec:ps} we review the dynamics of the clock and the dynamics of a generic fluid in a de Sitter background, highlighting the features that are relevant to understand the non-Gaussianities generated. In section \ref{sec:ng} we give the results for the calculation of the three-point function and a comparison of it with the recent Planck data.

\section{Effective field theory for inflation}\label{sec:eft}
~~~~~In this section we will review the basic elements taken from the effective field theory for inflation developed in \cite{diseft, diseftcon,eft1}. Although we will have the particular problem of warm inflation in mind for this study, we will develop the analysis using this approach to extract conclusions as general as possible in the context of dissipative single-clock inflation interacting with additional degrees of freedom defining a preferred frame. 
\subsection{Review of the formalism}
~~~~~In the simplest single field models, the EFT approach to inflation is based on the fact that there is a physical clock that defines a special time-slicing in which the clock is uniform and determines the end point of inflation. As noted in \cite{diseft,eft1} the field associated to this clock is related to the Goldstone boson breaking time translation symmetry, which is realized non-linearly. In the gauge where the clock is homogeneous, called the unitary gauge, its perturbations are encoded in the metric and the action is no longer time-diffeomorphism invariant. This way, the most general effective action can be written according to the symmetries available. Following St\"uckelberg's trick, one can also go to a gauge where perturbations are explicit by performing a time diffeomorphism $t\rightarrow t+\pi(x)$, where the field $\pi(x)$ transforms in a way such that the full diffeomorphism invariance is restored. The end result is an action for $\pi$ and metric perturbations which can be used to compute physical quantities, like curvature perturbations $\zeta=-H\pi+\ldots$ and its two and three point functions. This would be analogous to the renormalization gauge in gauge theories with spontaneous symmetry breaking.

In the case of an inflationary scenario in which more than one field is present, two distinctive cases are possible as we anticipated before. One case is multi-field inflation, in which the clock is coupled to a set of massless degrees of freedom that drive the clock fluctuations and generate peculiar features like iso-curvature perturbations, local shape of non-Gaussianities, etc. The characteristic feature of these ADOFs is the fact that they do not die away after horizon exit. In this case one defines the St\"uckelberg field $\tilde{\pi}$ (following the notation of \cite{diseft}) being the Goldstone boson associated with time translations, which drives inflation and the extra fields can affect the late time curvature perturbations. In this gauge the simple relation between $\zeta$ and $\tilde{\pi}$ is not straightforward. This kind of EFT of multi-field inflation was developed in \cite{multieft}. 

On the other hand, and this is the case of interest in this paper, one can have additional degrees of freedom with a contribution which decay at late times. In these cases, it is more natural to define a unitary gauge in which the physical clock that determines the end of inflation is uniform. Then one can introduce a St\"uckelberg field $\pi$ doing a time diffeomorphism. In this case the ADOF will contribute to the background and their tadpoles will depend on their vacuum expectation values. If the ADOF are represented by composite operators $\mathcal{O}$ then the relation with the gauge described in the previous paragraph is schematically $\tilde{\pi}\sim\pi+\delta\mathcal{O}$. In this gauge, measurable quantities like curvature fluctuations can be computed as $\zeta\sim -H\pi$, like in the single field EFT scenario. 

Within these models one can have dissipative theories of inflation like those studied in \cite{diseft} but one can also have quasi-single field inflationary models \cite{QSF}. The main difference between them is that in quasi-single field inflation correlations are driven by vacuum expectation values of extra field, with masses of the order of the Hubble. In dissipative inflation correlations are driven by expectation values for excited states through stochastic noise. This leads to differences in the non-Gaussianities. 
\subsection{Unitary gauge}
~~~~~In the unitary gauge there are no fluctuations of the preferred clock, the unit vector perpendicular to surfaces of constant time  ${t}$ takes the form  $ n_{\mu}=-\delta^{0}_{\mu} (-g^{00})^{-1/2}$, and the  extrinsic curvature of the surfaces is $K^{\mu}_{\nu}=\hat{g}^{\mu\rho}\nabla_{\rho}n_{\nu}$, where  $\hat{g}_{\mu\rho}=g_{\mu\rho}+n_{\mu}n_{\rho}$ is the induced spatial metric. The action is given by 
\begin{eqnarray}
\label{act1}
S&=&\frac{M_p^2}{2}\int d^4 x\sqrt{-g} R+\frac{1}{2}\int d^4 x \sqrt{-g}(\overline{p}-\overline{\rho}-(\overline{p}+\overline{\rho}) g^{00})\nonumber\\
&+&\frac{1}{2} \int d^4 x \sqrt{-g}\,M_2^4(t) (1+ g^{00})^2-\frac{1}{2}\int d^4 x \sqrt{-g}\,\overline{M}_1^3(t)\delta K_{\mu}^{\mu}(1+ g^{00})\nonumber\\
&-&\int d^4 x \sqrt{-g}\, f(t)\mathcal{O} +S_{\mathcal{O}}+\cdots,\label{action}
\end{eqnarray}where $M_{p}^2=(8\pi G_N)^{-1}$ is Planck's mass, $\mathcal{O}$ is a scalar composite operator associated with the dissipative degrees of freedom and $S_{\mathcal{O}}$ represents the action for this sector, which we do not need to specify. The dots stand for higher derivative corrections or other types of coupling to the ADOF. Bared quantities denote their unperturbed background values. Throughout the paper we assume the presence of an approximate shift symmetry, such that functions of time appearing in the action change very little in a Hubble time. This will be the basis of the generalized slow roll approximation, where not only $\epsilon=-\dot{H}/H^2\ll1$ and $\eta=\dot{\epsilon}/H\epsilon \ll 1$ but also $\epsilon_{g_i}=\dot{g}_i/g_iH\ll1$, with $g_i$ any parameter in the effective action. For the study of the squeezed limit of the three-point function the higher derivative terms are sub-leading.

In the unitary gauge where time diffeomorphisms are no longer a symmetry, the metric can be written using the ADM decomposition as 
\beq
ds^2=-N^2dt^2+ a^2(t)\delta_{ij} e^{2\zeta} (dx^i+N^i dt) (dx^j+N^j dt),
\eeq
where $\zeta$ is the curvature perturbation and we ignore tensor perturbations which are sub-leading in slow-roll approximation. After performing St\"uckelberg trick one introduces the preferred clock perturbation $\pi$ through $t\rightarrow t+\pi$. Now the metric can be decomposed in the same way setting $\zeta=0$ in the expression above. Following \cite{maldacena}, the gauge invariant curvature perturbation $\zeta$ will be, to lowest order in $\pi$, 
\beq
\zeta=-H\pi + \mathcal{O}(\pi^2,\pi\partial\pi,(\partial\pi)^2,\epsilon,\text{ etc})
\eeq
We will be interested in the effect on non-Gaussianities of fluid-like excitations of the ADOF at zeroth order in the slow-roll parameters; therefore we will not care about these corrections. 
\subsection{Dynamics of perturbations in $\pi$-gauge}
~~~~~In this section we will derive the action describing the dynamics of $\pi$. We will neglect gravity fluctuations and any term sub-leading in slow-roll and we will concentrate on scalar modes only. Neglecting gravity perturbations can be understood in terms of an analogous equivalence theorem in gauge theories, where the Goldstone boson decouples from the transverse polarization of the gauge bosons \cite{eft1}. The effects of the ADOF can be described writing the composite operator as composed of three contributions: the background value $\bar\calo(t)$, the response $\delta\calo_R (t,{\bf x})$ due to the coupling to $\pi$  and a stochastic component $\delta\calo_{\cal S}(t,{\bf x})$ such that
\beq\label{calotxsr}
\mathcal{O} (t,{\bf x})= \overline{\mathcal{O}} (t)+ \delta \mathcal{O}_{\mathcal{S}}(t,{\bf x}) + \delta\mathcal{O}_R (t,{\bf x}).
\eeq
The derivation of this fact from first principles is rather involved \cite{calzetta} but an effective description of each contribution can be guessed easily.

We will first consider the tadpoles describing the background evolution. For the coupling in equation \ref{action} the requirement of having an inflationary epoch gives the equations 
\begin{eqnarray}
&& 3H^2 M_p^2=\bar{\rho}+\bar{\rho}_{\calo}+f(t)\bar\calo, \label{beqmet}\\
&&\dot{\bar\rho}+3H(\bar\rho+\bar p)+\dot{f}\bar{\calo}=0,\label{beqpi}\\
&&\dot{\bar\rho}_{\calo}+3H(\bar\rho_{\calo}+\bar p_{\calo})+f\dot{\bar{\calo}}=0,\label{bacO}
\end{eqnarray} where we used that the background stress tensor for the ADOF obtained from $S_{\calo}$ takes the perfect fluid form
\beq\label{tmunuO}
{\bar T}_{\mu\nu}^{\calo} = (\bar \rho_{\calo}+\bar p_{\calo}){\bar n}_\mu {\bar n}_\nu + \bar g_{\mu\nu} \bar p_{\calo}.
\eeq 

If we introduce $\pi$ following the St\"uckelberg trick, and expand the action in powers of $\pi$, including interactions with ADOF  $\sim\pi\delta\calo$, then one gets 
\begin{eqnarray}\label{l2pi}
S=\int d^4x a^3\frac{N_c}{2}\left\{\dot{\pi}^2-c_s^2\frac{(\partial_i\pi)^2}{a^2}\right\}+S_{\calo}-\int d^4x a^3\frac{\dot{f}}{2}\pi \left(\delta\calo^s_R + \delta\calo_{\cal S}\right)+\cdots,
\end{eqnarray}
where
 \beq \label{ncs} c_s^2=\frac{(\overline{p}+\overline{\rho}+H\overline{M}_1^3)}{(\overline{p}+\overline{\rho}+4M_2^4)},
 \eeq
and $N_c=(\overline{p}+\overline{\rho}+4M_2^4)=(\overline{p}+\overline{\rho}+H\overline{M}_1^3)/c_s^2$. The dots represent terms of higher order in $\pi$ or higher order in slow-roll parameters. Non-Gaussianities arising from those terms have been thoroughly studied previously using a local approximation of the $\delta\calo$ response \cite{diseft}. This approximation means that we are thinking of an inflaton dissipating energy into an environment of ADOF at rest which later gets diluted by Hubble expansion or coupling with extra sectors. A realization of this idea is given by, for example, trapped inflation \cite{trapped} with an extra sector making the particles decay faster than Hubble. 

On the other hand, in this paper we will focus on how time non-localities in the dynamics of the ADOF (like the one that appears when there are sound waves in the environment which decay in a time-scale of a Hubble) affects the squeezed limit, the consistency condition and generally the shape of non-Gaussianities. The contributions previously studied in \cite{diseftcon} satisfy the consistency condition and are slow-roll suppressed in the squeezed limit and we will not consider them here. 

Before computing the particular features of this theory, two more steps are needed. The first one is specifying the statistics of the stochastic component of $\calo$, which we write as 
\beq
\label{twop1}
\langle \delta\calo_{\cal S}(t,{\bf k})\delta\calo_{\cal S}(t',{\bf q})\rangle \simeq \frac{\nu_{\calo}(t)}{\sqrt{-g}}\delta(t-t') (2\pi)^3 \delta^{(3)}({\bf q+\bf k}),
\eeq
which is written in terms of spatial Fourier decomposition in comoving coordinates and $\nu_{\calo}(t)$ is the noise kernel. Due to the underlying shift symmetry the noise kernel should be time independent to lowest order in slow-roll.  

The last piece needed is to get an expression for the response part of $\calo$ in terms of $\pi$. This was argued in \cite{diseft, diseftcon} to be of the form 
\beq
\label{respo0}
\bar\calo+ \delta \calo^u_R \simeq  \Lambda_\calo(t)~F\left[\sqrt{-g^{00}}, K^\mu_\mu, t \right],
\eeq
in the unitary gauge. Going to the $\pi$-gauge and retaining only the lowest order in $\pi$ one arrives to the expression 
\beq\label{res1}
\delta\calo_R(t)\simeq V_{\calo}(t)\dot{\pi}
\eeq
The purpose of this paper is to study the effects of having corrections to this response coming from the presence of fluid-like excitations in the ADOF, presenting sound waves which mediate interactions non-locally in time. In these cases the response will be of the form $\delta\calo_R \sim \int dt' G(t,t')\pi(t')$ with $G(t,t')\neq0$ for $|t-t'|\lesssim H^{-1}$. This will be done in the next section. Of course while doing this one looses generality because the parameter space describing the most general non-local interaction is much bigger than the local one. Nevertheless, since the kernels of this non-localities are given by the retarded Green functions of the ADOF we will study the behavior of these functions for a fluid in de Sitter.

In principle, this parameter $V_{\calo}$ is arbitrary, but if the theory is, for example, in thermal equilibrium at temperature $T$ (warm inflation) there is a specific relation connecting $V_{\calo}(t)$, $\nu_{\calo}(t)$ and $T$, namely the fluctuation-dissipation theorem (FDT) in that scenario.

At this point one can start computing relevant quantities. The procedure is outlined for example in \cite{diseft, diseftcon,eft1}. The first thing to do is getting the equations of motion for $\pi$ from the action given in equation \eqref{l2pi}, which will involve $\calo_R$ and $\delta\calo_{\mathcal{S}}$. Then, one uses the expression of $\calo_R$ in terms of $\pi$ and solve the equation obtaining $\pi\propto \delta\calo_{\mathcal{S}}$. Finally, thanks to the gauge we are using, one can easily obtain $\zeta \sim -H\pi \propto \delta\calo_{\mathcal{S}}$ and compute its correlation functions using the statistics of the stochastic source.  
\subsubsection{Matching warm inflation: Background and Power Spectrum}
~~~~~In the case of warm inflation \cite{warm,moss,moss2} the matching with the EFT parameters is straightforward. As explained in the Introduction, in this model the ADOF is radiation in thermal equilibrium at a temperature $T$ such that $T\ll V^{1/4}$ and $T\gg H$, where $V$ is the inflaton potential. The coupling modifies the dynamics of the inflaton perturbations but in this temperature range its energy-momentum density is subdominant with respect to the potential energy $V\sim M^2_pH^2$, which drives the exponential expansion. Moreover, in this regime thermal fluctuations dominate over vacuum fluctuations.

The inflaton sector is the same as in the canonical single-field inflation scenario. One has a scalar field $\varphi$ with a time dependent expectation value $\overline{\varphi}(t)$ that drives inflation, with perturbations around this background $\overline{\varphi}(t)+\delta\varphi({\bf x},t)$. In the simplest model, it has the canonical action $S=\int \sqrt{g} [(\partial \varphi)^2 -V(\varphi)]$ which in unitary gauge is 
\beq
S_{\varphi}=\frac{1}{2}\int \sqrt{g}[-(\dot{\overline{\varphi}})^2g^{00}-2V(\overline{\varphi})],
\eeq
giving the expected contribution to the tadpoles. Perturbing the background and going to the $\pi$-gauge gives the identification $\pi=\delta\varphi/\dot{\overline{\varphi}}$, as usual \cite{diseft,eft1}. 

For the model considered in \cite{moss,moss2}, the interaction of the fluid with the inflaton is given by the rate of energy exchange between them due to particle creation, which is written as:
\beq\label{entransf}
\nabla_\mu T^{\mu}_{~\nu}=-\gamma u^{\mu}\partial_\mu\varphi\partial_\nu\varphi,
\eeq where $\gamma$ is a parameter given by the underlying microscopic theory for this energy transfer, $T^{\mu\nu}$ is the inflaton stress-energy tensor and $u$ is the 4-velocity of the fluid, which in the background coincides with $n^{\mu}$ and will give the second preferred frame when perturbations are included. 

The equations of motion for the background (or tadpoles in the language of the field theory) in the slow-roll regime are 
\begin{eqnarray}
&&(\gamma+3H)\dot{\varphi}+V_{\varphi}\simeq 0,\\
&&4H\rho_r\simeq\gamma \dot{\varphi}^2,\\
&&3H^2M_p^2\simeq V. 
\end{eqnarray}
Again, these equations include the assumption that, even if $T\gg H$, since $T\ll V^{1/4}$ the radiation energy density is small compared to the inflaton and then the Hubble constant evolves according to $V(\varphi)$. 

Using equation \eqref{entransf} one can write an action for $\pi$ and identify the different quantities of the effective field theory. In the simplest setup of warm inflation, $c_2=1$, $\dot{f}=1$ and the normalization is $N_c=\dot{\varphi}^2$. This gives the equation of motion for inflaton perturbations to lowest order 
\beq\label{wipiaLO0}
\ddot{\pi}+3H\dot{\pi}-\frac{\partial^2}{a^{2}}\pi+\frac{1}{N_c}\delta\calo_R^{(1)}(t,{\bf x})=-\frac{1}{N_c}\delta \calo_{\cal S}(t,{\bf x}),
\eeq  
and in order to match the dynamics with the warm inflation the response should be $\delta\calo_R(t)\simeq V_{\calo}(t)\dot{\pi}$ with $V_\calo \simeq\gamma N_c$. This gives the following EOM
\beq\label{eqpilin}
\ddot{\pi}+(3H+\gamma)\dot{\pi}-\frac{\partial^2}{a^{2}}\pi=-\frac{1}{N_c}\delta \calo_{\cal S}(t,{\bf x}).
\eeq
Besides, the FDT gives the noise kernel in terms of the dissipation coefficient, namely $\nu_\calo=2\gamma T$. Using this description of the clock one can compute the two point function to get the curvature power spectrum using $\zeta \sim -H\pi$ and relate this with observable curvature anisotropies. We review the calculation in section \ref{sec:ps}. 
\subsection{Interactions}
\subsubsection{Shift Symmetry}
~~~~~ In \cite{diseft} interactions are introduced in several ways. The first one is due to the approximate nature of the shift symmetry. For example the parameters in the Lagrangian describing the de Sitter background are not exactly time independent. This induce non-linear terms in the equations of motion when expanded in terms of $\pi$. These are proportional to the slow-roll parameter. They are bound to be small but otherwise unspecified by the background. These terms are relevant in the power spectrum since they parameterize the breaking of scale invariance through the tilt $n_s$. They also generate an apparent local contribution to the three-point function which exactly saturates the consistency condition as was shown in \cite{diseftcon}. These terms are present and unchanged in the set-up considered here. Therefore any additional contribution to the squeeze limit of the three-point function would result in a violation of the consistency condition.
\subsubsection{Response}
~~~~~The second source of non-linearity is due to corrections to the response of the ADOF. Even tough they might be constrained in each particular scenario of inflation, they are new parameters from an EFT perspective because they do not influence the evolution of the background space-time. In this case their behavior is different than in \cite{diseft} because of the presence of corrections which are non-local in time   
\beq
\delta\calo_R~\supset \int dt' G_R(t,t') \partial_\mu^{(n)}\pi(t')\partial^{(m)}_\nu\pi(t)+\ldots,
\eeq
where $G_R(t,t')$ is the retarded Green function of the ADOF and $n,m$ is the number of derivatives. (In the case of warm inflation that we have in mind, this kind of non-linearities will be generated for example by a temperature dependence of the dissipation coefficient and they will be multiplied by an arbitrary factor of $\partial_T\gamma$, since 
\beq
\partial_T\gamma\delta T\dot{\pi}\sim\partial_T\gamma T  \dot{\pi}(t) \int G_R(t,t') \partial^2\pi(t').
\eeq
 As opposed to $\gamma$ itself, the strength of this coupling is arbitrary, it is not fixed either by the power spectrum nor the background symmetries -although a stability analysis of warm inflation shows that this parameter cannot be too large $\partial_T\gamma<4\gamma/T$, see \cite{wistability}).
 
In the paragraph above we were considering generic corrections to the response. Nevertheless, there is a specific type of corrections due to the non-linear realization of coordinate invariance, and they are bounded by the background parameters like $\gamma, H$, etc. The idea is the following: if one has a term $\gamma\dot{\pi}$ in the equation of motion clearly this is not diffeomorphism invariant. Therefore at the nonlinear level this must be changed by $\gamma w^\mu \partial_\mu \pi$, with $w$ some time-like four vector that reduces to $\delta^\mu_0$ to lowest order in fluctuations. For a single preferred frame then $n^\mu$ is the only choice. In the case studied here since we have two preferred frames we could either use $n^\mu$ or $u^\mu$, the fluid four vector defining its frame. The one generated by $n^\mu$ was studied before and relates the linear response, characterized by $\gamma$, with the nonlinear response giving $|f_{\rm NL}|\sim \gamma/Hc_s^2$ so in this paper we will focus on the one generated by $u^\mu$ that shares the same property of relating linear with nonlinear phenomena.

Instead of going through the action, here we will work directly with the energy transfer between the clock and the ADOF, which we will treat as a generic fluid. The approach developed in \cite{diseft} can be followed, but the fact that another time-like vector operator is present besides $n^\mu$ increase the number of possible terms in the EFT respecting the underlying symmetries and the background\footnote{Nevertheless, this does not affect the construction of the Goldstone boson $\pi$ and the unitary gauge frame since what the clock fixes is the time at which inflation ends and we will see that for a typical fluid its perturbations decay after the modes exit the sound horizon and therefore these fluid-like ADOF become unobservable. In other words, at the time inflation ends $u^\mu \simeq n^\mu $ and the definition of the clock is unambiguous since there is only one relevant frame. If this were not the case, one would be facing a multifield-like model of inflation with observable degrees of freedom.}. To get the required dissipation term for the background one needs a flux of energy of the form $Q_{(0)}^\mu =-N_c\gamma \delta^\mu_0$. To non-linearly realize diffeomorphism invariance in a way that gives the desired background the first guess is $Q^\mu = -N_c \gamma n^\mu$. Since now we have a new time-like vector $u^\mu$ the most general realization is 
\beq
\overline{Q}^\mu+\delta Q^\mu = -N_c\gamma \left( n^\mu F_n \left[\sqrt{-g^{00}},n\cdot u, \rho,K^\mu_\mu,t\right] + u^\mu F_u \left[\sqrt{-g^{00}},n\cdot u,\rho,K^\mu_\mu,t\right]\right),
\eeq    
where $\rho$ is the energy density of the fluid and $F_n [1, \overline{\rho},3H,t] +  F_u [1,\overline{\rho},3H,t]=1$. One can also define the current $J^\mu = \rho u^\mu$ and write everything in terms of $J^{\mu}$ instead since the scalar $J^2$ is not fixed. The explicit time dependence will be suppressed by the shift symmetry and will be sub-leading in the slow-roll parameters. For example, in order to reproduce perturbations in the warm inflation scenario as the ones studied in \cite{moss} one has to take $F_u=0$ and $F_n=\sqrt{-g^{00}}n\cdot u + \ldots$ where the dots represent subleading corrections, for example temperature dependence of $\gamma$, i.e. $\partial_\rho F_n \neq 0$. To obtain the equation of motion for the clock the $\mu=0$ term is enough but the other components are necessary to specify the fluid dynamics. The leading correction to the clock equation due to the second term is proportional to $\delta u^0\sim \calo(\pi^2)$ and therefore the leading non-Gaussianities come from the first term. 

To lowest nonlinear order they will appear as $$\gamma_n n^\mu \partial_\mu \pi+\gamma_u u^\mu \partial_\mu \pi$$ in the equation of motion for $\pi$, where $\gamma_u+\gamma_n=\gamma$ to match the correct linear result. These will give different contributions to the three-point function as we will see below.

Once these functions are specified one can solve the equations of motions of the fluid using energy conservation. One can use the most general stress-energy tensor for a fluid parametrized by a certain set of coefficients \cite{fluidos}. Even if one has a perfect fluid those ``irrelevant" terms can appear from an effective field theory approach due to interactions mediated by other degrees of freedom or even self interactions of the fluid. Anyway, here we will consider the usual case of a perfect fluid
\beq\label{tmunur}
{T_r}_{\mu\nu} = (\rho+p)u_\mu u_\nu +  p g_{\mu\nu},
\eeq 
 parameterized by an equation of state, although the generalization is direct. The presence of sound waves in the ADOF makes the self-ineractions behave non-local in time and therefore one can think of this procedure as a physical motivation of the non-localities and eliminate $u$ and $\rho$ from the start. 

(For the particular case of warm inflation, one could assume an energy flux of the form $\gamma u^\mu \partial_\mu\varphi \partial_\nu \varphi$. This is assumed without much justification in \cite{moss} and it is not clear whether the most natural velocity to appear is $u^\mu$ or $n^\mu$ since we have now two time-like vectors at hand with the same background value. To avoid this, one could assume a more general case of the form
\beq\label{eq:energyflux}
\nabla_\mu T^{\mu}_{~\nu}=-\gamma_u u^{\mu}\partial_\mu\varphi\partial_\nu\varphi-\gamma_nn^{\mu}\partial_\mu\varphi\partial_\nu\varphi,
\eeq
and call $\gamma\equiv\gamma_u+\gamma_n$ like before the dissipation coefficient that appears in the background and power spectrum. Given that the system is closed and overall energy conservation holds, equation \eqref{entransf} is also useful to get an equation for the evolution of perturbations in the fluid radiation since the energy lost by $\pi$ should equal the energy gained by the fluid 
\beq\label{entransfrad}
\nabla_\mu {T_r}^{\mu}_{~\nu}=\gamma_u u^{\mu}\partial_\mu\varphi\partial_\nu\varphi+\gamma_nn^{\mu}\partial_\mu\varphi\partial_\nu\varphi,
\eeq 
where $T_r$ is the energy-momentum tensor of the radiation which has the form shown in equation \eqref{tmunuO} with $n^{\mu}\rightarrow u^{\mu}$. Using this expression, equation \eqref{entransfrad} gives the equation of motion for the ADOF.)

\subsubsection{Stochastic Component}
~~~~~The last type of non-linear terms that appear in the EOM are due to corrections in the statistics of the stochastic component of the ADOF response. These can be generated as corrections to Gaussian behavior adding non zero three-point function of $\delta\calo_{\mathcal{S}}$ and this was studied in \cite{diseft}\footnote{There are also terms coming from $\nu_\calo\neq0$ but we include them in the group of shift symmetry violating terms.}. These corrections appear in the same way in the set-up considered in this paper and the analysis is unchanged.

 On the other hand, terms can be generated letting the two-point function have an arbitrary dependence on scalar quantities build up from $n^\mu$ and $J^\mu=\rho u^\mu$. This can happen if the underlying physics constraint the noise to have a specific dependence on other variables, as in the case of the FDT \cite{moss,moss2}. We can write:
\beq
\lb \calo_S^2 \rb \sim\nu_\calo F_\nu[\rho, \sqrt{-g^{00}},u^\mu n_\mu,K_\mu^\mu,t],
\eeq
since corrections in the statistics can be taken into account adding a $\pi$ dependence of the kernel \cite{diseftcon}. This function must satisfy $F_\nu[\overline{\rho},1,1,3H,t]=1$. Here we will study only the one coming from the dependence on the first argument. In principle this is not fixed by symmetries or the background but if the FDT appears it forces a specific value.

\section{Power spectrum and correlations}\label{sec:ps}
~~~~~In this section we will review the results for the power spectrum $\lb \pi\pi\rb$ of dissipative single-clock inflation, computed in \cite{diseft}, mainly to write some results that will be useful later. We will also compute the solution and Green function of the EOM describing the dynamics of an ideal fluid with a generic equation of state. This will allow us to characterize the general behavior of non-local interactions mediated by sound waves. For example, fluid-like perturbations decay after horizon-exit having strong implications when computing the three-point function $\lb \zeta^3\rb$ that characterizes non-Gaussianities. If ADOF fluctuations remain after horizon exit, that would mean that they behave as almost massless degrees of freedom like in the multi-field inflation case \cite{multieft}. On the other hand, if perturbations decay then the dissipative single-clock EFT of \cite{diseft, diseftcon} is more appropriate.
 \subsection{Power spectrum}
 ~~~~~The action \eqref{l2pi} gives the following equation for the clock perturbation $\pi$ at linear level including the effects of dissipation and noise   
\beq\label{piaLO}
\ddot{\pi}+(3H+\gamma)\dot{\pi}-\frac{c_s^2}{a^2}\partial^2\pi=-\frac{1}{N_c}\delta \calo_{\cal S}(t,{\bf x}),
\eeq 
where we omit factors such as $\dot{f}$ since they could be included in the noise two-point function without affecting the results of this section. Looking at the expression above, the power spectrum derived from it will depend only on the combination $\gamma=\gamma_u+\gamma_n$ characterizing the background. The solution to this equation was found in \cite{moss,diseft} and it is given by\footnote{Of course, the full solution contains an homogeneous component with the initial conditions specifying the initial state, for example the Bunch-Davies vacuum. Nevertheless, this component decays in the $\gamma\gg H$ limit when $t_0\rightarrow-\infty$ and therefore its effect is negligible as shown in \cite{diseft}.}  
\beq \label{pi1}
\pi_1(\eta,{\bf k})=\frac{  k c_s}{N_cH^2}\int^{\eta}_{\eta_0} d\eta' G_{\gamma}(k c_s|\eta|,k c_s|\eta'|)\frac{\delta{ \calo}_S}{(kc_s|\eta'|)^2},
\eeq
where $\pi({\bf k},t)$ is the Fourier transform with comoving wave-vector ${\bf k}$, $\eta$ is the conformal time $\eta= \int dt/a\simeq-(aH)^{-1}$, $\eta_0$ is an early enough initial time, and the clock's Green function is given by:
\beq
\label{greengamma}
G_{\gamma}(z,z')=\frac{\pi}{2} z \left(\frac{z}{z'}\right)^{\nu-1} \left[Y_\nu(z) J_\nu (z')-J_\nu (z) Y_\nu(z')\right],
\eeq 
with $\nu=\frac{3}{2}+\frac{\gamma}{2 H}$, $z=-k c_s\eta$, and  $z'=-k c_s\eta'$. We also define the Green function that includes the factor coming from the change of variable from $t$ to $\eta$ as $g_\gamma(z,z')=G_\gamma(z,z')/{z'}^2$. 
\begin{figure}[t!]
\begin{center}
\subfigure[]{
\includegraphics[scale=0.38]{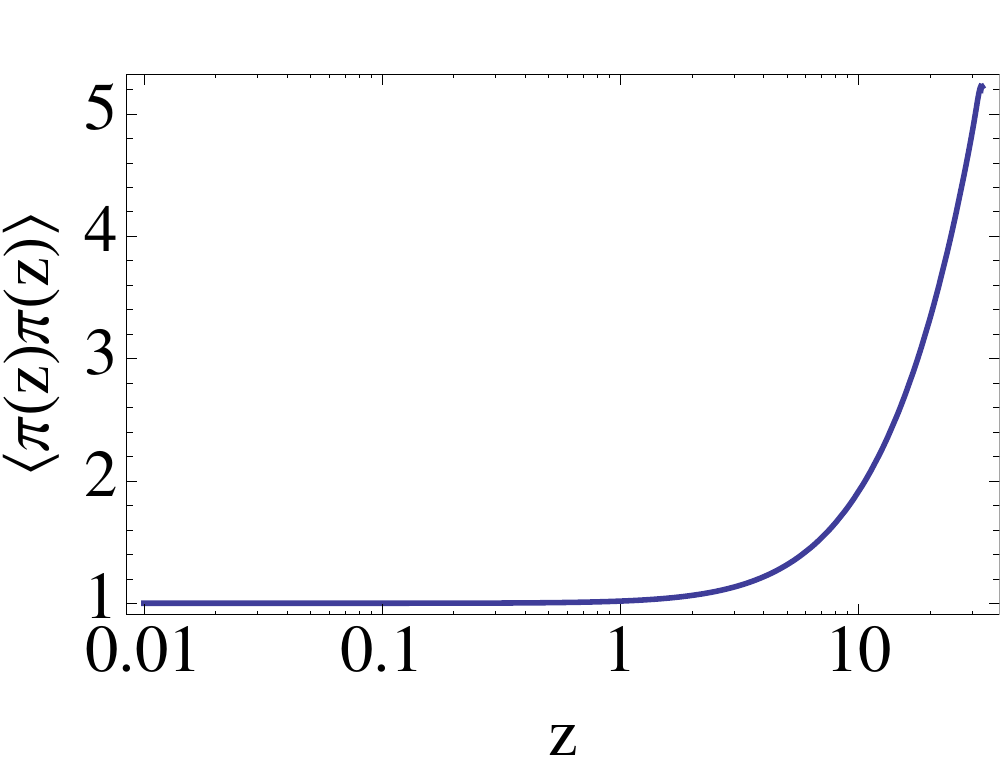}}
\subfigure[]{
\includegraphics[scale=0.4]{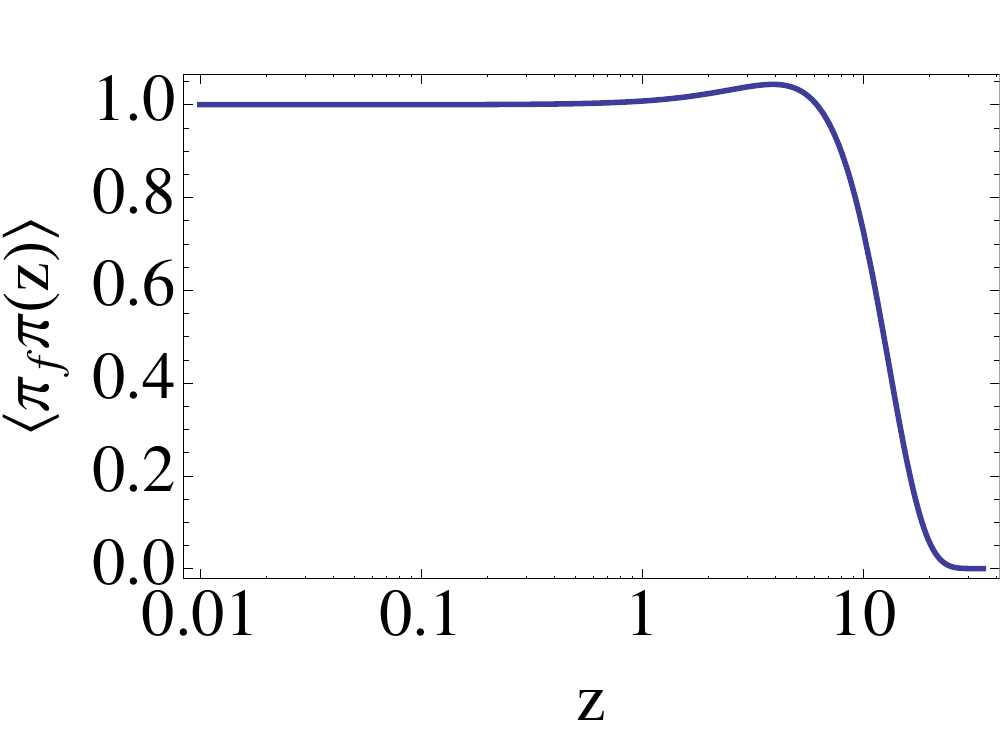}}\\
\subfigure[]{
\includegraphics[scale=0.35]{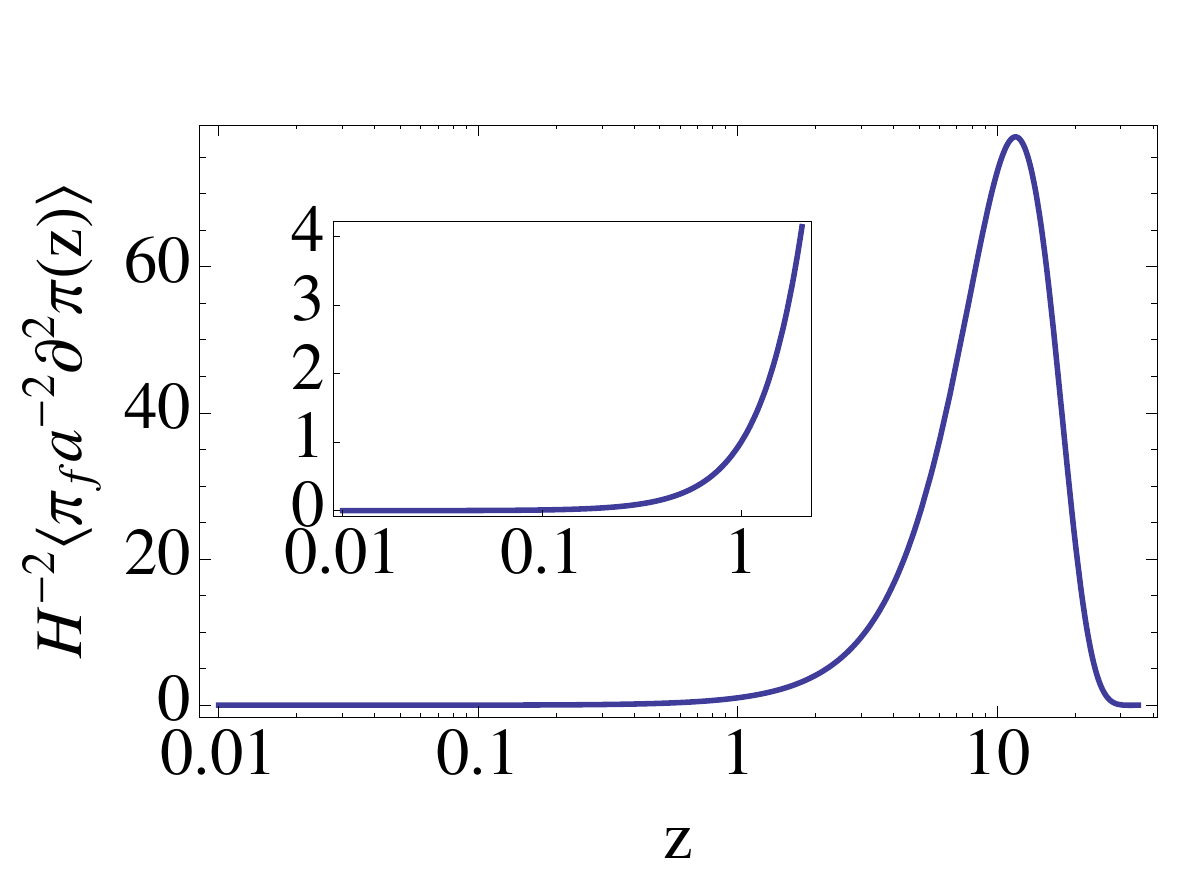}}
\subfigure[]{
\includegraphics[scale=0.4]{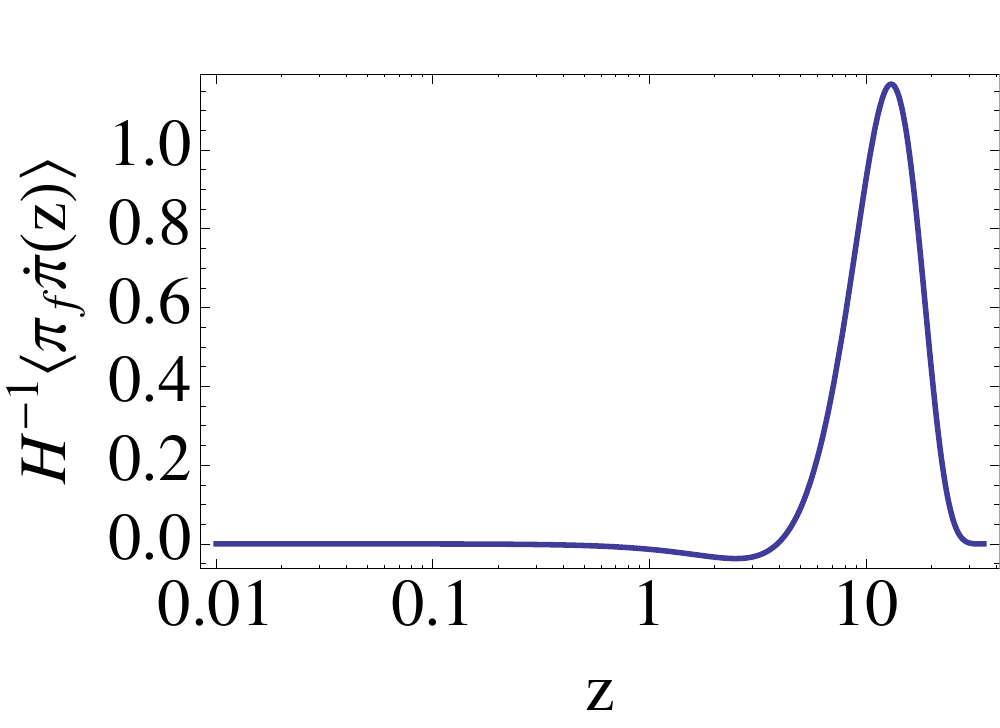}}
\caption{\label{fig:pi1} \small\it Plot for a dissipation coefficient $\gamma=50 H$ of the correlations (a) $\lb \pi(z)\pi(z)\rb$ (b) $\lb \pi(z)\pi_f\rb$ (c) $\lb a^{-2}\partial^2 \pi \pi_f\rb$ (d) $\lb \dot{\pi} \pi_f\rb$. All of them are normalized with respect to $\lb \pi_f\pi_f\rb$ for $\gamma=50H$. The behavior for other dissipation coefficient is similar.}
\end{center}
\end{figure}
Using the explicit expression for the two point function of the stochastic component \eqref{twop1}, the power spectrum of curvature perturbations $\zeta \sim H\pi$ at late times, defined as
\beq
\lb \zeta_{\mathbf{k}_1}\zeta_{\mathbf{k}_2}\rb=(2\pi)^3\delta(\mathbf{k}_1+\mathbf{k}_2)P_\zeta(k).
\eeq 
can be computed and it is given by
\bea \nonumber
  P_{\zeta}(k)&=&\lim_{z\rightarrow0,~z_0\rightarrow\infty}~\frac{\nu_{\calo}H^2 }{N_c^2(kc_s)^3}\int^{z_0}_z dz' G^2_{\gamma}(z,z')\\
  &=&\frac{\nu_{\calo}^\star H^2_\star }{k^3{N_c^\star}^2{c_s^\star}^3}\frac{16^{\frac{\gamma_\star}{H_\star} } (\frac{\gamma_\star}{H_\star} +1)^3
 \Gamma \left(\frac{\gamma_\star +H_\star}{2H_\star}\right)^4}{\pi  \Gamma ( \frac{2\gamma_\star}{H_\star} +4)}\simeq\frac{\nu_{\calo}^\star}{k^3} \sqrt{\frac{\pi H_\star}{\gamma_\star}} \frac{H_\star^2}{2c_s^\star\left({c^\star_s}N_c^\star\right)^2}.
\ea
 All the quantities are evaluated at freeze-out time $\omega_\star=c_s k /a(t_{\star})\simeq \sqrt{\gamma_\star H_\star}$ or equivalently $c_s k |\eta_\star| \simeq  \sqrt{\gamma_\star /H_\star}$. For example, with this expression one can obtain the tilt $n_s-1$ due to slow roll corrections and the result is given in \cite{diseftcon}. The right hand side of the second line is valid in the strong dissipation regime $\gamma\gg H$. The time dependence of the $\zeta$ power spectrum is plotted in figure \ref{fig:pi1}(a) and it can be seen how $\pi$ becomes constant after the freeze-out time defined above. As explained in \cite{diseft} this is because the equilibration time defined by the dissipative response $\tau_{eq} = \gamma/\omega^2$ gets larger than the expansion rate given by the Hubble $H^{-1}$ at freeze-out and therefore the noise-induced curvature perturbation decouples from the expansion and remains constant. 

In the following sections we will see how inhomogeneities of the clock $\pi$ can drive fluid perturbations and in turn produce non-Gaussianities. These are driven by correlations like $\lb a^{-2}\partial^2 \pi \pi_f\rb$, $\lb \dot{\pi} \pi_f\rb$, $\lb \ddot{\pi}\pi_f\rb$, where $\pi_f$ is the clock perturbation at late times (after freeze out). These can all be derived from $\lb \pi_{\mathbf{k}}(z)\pi_{\mathbf{k}f}\rb$, which is given by 
\bea \label{pipif}
\frac{\lb \pi_{\mathbf{k}}(z)\pi_{\mathbf{k}f}\rb}{\lb \pi_{\mathbf{k}f}\pi_{\mathbf{k}f}\rb} &=& \frac{\pi  2^{-7 \nu } z^{\nu } }{\Gamma(\nu)\Gamma (\nu -1)}\bigg[\frac{128 \pi  \Gamma (4 \nu -2) J_{\nu }(z)}{\Gamma (\nu )} \nn
&&~\bigg(\frac{z^{3-2 \nu } \, _2F_3\left(\frac{1}{2},\frac{3-2\nu}{2} ;1-\nu ,\frac{5-2\nu}{2} ,\nu +1;-z^2\right)}{3 \pi  \nu -2 \pi  \nu ^2}\nn
&&~+\frac{z^3 \Gamma (-\nu ) \, _2F_3\left(\frac{3}{2},\nu +\frac{1}{2};\frac{5}{2},\nu +1,2 \nu +1;-z^2\right)}{3 \sqrt{\pi } \Gamma \left(\frac{1}{2}-\nu \right) \Gamma (2 \nu +1)}-\frac{\Gamma \left(\frac{3}{2}-\nu \right)}{4 \Gamma (2-\nu ) \Gamma \left(2 \nu -\frac{1}{2}\right)}\bigg)\nn
&&~ - 64^{\nu } Y_{\nu }(z) \bigg(\Gamma (\nu -1)-z^3 \Gamma ((2\nu -1)/2)\Gamma((2\nu +1)/2) \Gamma ((4 \nu -1)/2)\nn
&&~\times \, _2\tilde{F}_3\left(\frac{3}{2},\nu +\frac{1}{2};\frac{5}{2},\nu +1,2 \nu +1;-z^2\right)\bigg)\bigg],
\ea
where the $(2\pi)^3\delta(\sum\mathbf{k})$ factors are omitted\footnote{This expression presents apparent singularities for integer or half-integer $\nu$ but different contributions cancel and give a continuous behavior.}. To understand their behavior, relevant correlations are shown in figure \ref{fig:pi1}(b), (c) and (d). 

First of all, one can see in figure \ref{fig:pi1}(b) that $\lb \pi_{\mathbf{k}}(t)\pi_{{\mathbf{k}}f}\rb$ is zero for early times and becomes a constant near $z(t)\simeq z_\star$. As explained before, this is because when freeze-out occurs the ``memory" of the system quantified by $\tau_{\rm eq}$ gets larger than the expansion rate and therefore perturbations do not decay with time and in turn survive and affect $\pi_f$ giving a constant non-vanishing correlation. The following property can be checked from the formula \eqref{pipif}. For $\gamma \gg H$, the right-hand side of it only depends on the ratio $z/z_\star$ and therefore 
\beq
\lb \pi_{\mathbf{k}}(z)\pi_{\mathbf{k}f}\rb\simeq \lb \pi_{\mathbf{k}f}\pi_{\mathbf{k}f}\rb \mathcal{F}(z/z_\star).
\eeq
 This means that the normalized shapes shown in figure \ref{fig:pi1} are valid for arbitrary, but strong, dissipation after the appropriate rescaling.

On the other hand, the correlation $\lb a^{-2}\partial^2 \pi_{\mathbf{k}}(t) \pi_{\mathbf{k}f}\rb$ measures the inhomogeneities in the clock. Before freeze-out the mode is not correlated to $\pi_{\mathbf{k}f}$ and therefore this correlation vanishes. After freeze-out $\lb \pi_{\mathbf{k}} \pi_{\mathbf{k}f}\rb$ is constant and the mode's physical length scale $\lambda_{\rm pays}$ is very small since it scales as $\gamma^{-1/2}$, giving a parametrically large correlation scaling as $\gamma$ at freeze-out (with respect to $\lb \pi_{\mathbf{k}}^2\rb$) and decaying as $\lambda_{\rm phys}^{-2}$. At later times when the mode exits the Hubble horizon at $\lambda_{\rm phys} \simeq H^{-1}$ the mode becomes constant over a Hubble patch and the correlation decays. This behavior can be seen in figure \ref{fig:pi1}(c).

Finally, correlations between $\pi_f$ and $\dot{\pi}$ or $\ddot{\pi}$ are peaked near freeze-out $z_\star$ for the same reason but decay more quickly and their strength can be orders of magnitudes below $\lb a^{-2}\partial^2 \pi \pi_f\rb$. The reason is that the time scale of variation of $\pi$ around freeze-out is fixed by the equilibration time $\tau_{\rm eq}$ which at freeze-out is given by $H^{-1}$ so there is no enhancement coming from factors of $\gamma$ as it happens for $\lb a^{-2}\partial^2 \pi_{\mathbf{k}}(t) \pi_{\mathbf{k}f}\rb$. One can check comparing figure \ref{fig:pi1}(c) and \ref{fig:pi1}(d) that indeed the correlation with time derivatives is suppressed by a factor of the order of $\gamma/H$ with respect to spatial derivatives. 

\subsection{Fluid Perturbations}
~~~~~In this section we will study the dynamics of fluid perturbations which will source non-Gaussianities of the three-point function through a non-vanishing expectation value for a time-like four vector $u^\mu$. It was shown in \cite{diseft,eft1} that in dissipative single-clock inflation metric perturbations decouple from the clock perturbations for energies $E\gg\sqrt{\epsilon}H$, which in the strong dissipation regime is clearly satisfied $E\sim \sqrt{\gamma H}$. Therefore, we can get a sense of the physics in these models ignoring gravity fluctuations overall from the start if we are working to lowest order in slow-roll. Then, the equation of motion for the fluid will be given by its conservation law in a de Sitter background
\beq
ds^2\simeq-dt^2+ a^2(t)\delta_{ij} dx^i dx^j.
\eeq
with a scale factor with Hubble parameter $H$ approximately constant over time, with slow-roll suppressed time dependence. 

We will only consider scalar perturbations in the fluid since vector perturbations decay faster than the scalar ones. As we said, we will also neglect anisotropic stress in this work or any term in the fluid stress-tensor other than the perfect fluid ones. A more systematic treatment using the most general theory for a fluid could be done using for example the formalism from \cite{fluidos}. 

The background values of the fluid degrees of freedom $\overline{\rho},\overline{p}$ are constant in time, and in particular $\overline{u}^i=0$. From now on we will not write the line over background values when it is unambiguous. Following the notation of \cite{weinberg} for example, the perturbations of the radiation energy-momentum tensor \eqref{tmunur} can be expanded in a standard way as
\bea\label{ap:pert}
\delta T^0_{~0}&=&-\delta\rho,\\
\delta T^0_{~i}&=&(\rho+p) u_i ,\\
T^i_{~j}&=&\delta_{ij}\delta p+\Pi_{ij},
\ea
where $\Pi_{ij}$ represents all the terms coming from deviations from the perfect fluid which we are neglecting in this work. This might include dissipative effects in the fluid itself.

The equation of motion for the fluid follows most easily from energy conservation $\nabla_{\mu} T^{\mu}_{~\nu} = Q_{\nu}$, where $T^{\mu\nu}$ is the radiation stress energy tensor and $Q_{\nu}$ 
 is the energy flux between the clock and the fluid. The background satisfies this equation with the clock and fluid frame coinciding $u^\mu=n^\mu\simeq\delta_0^\mu$ and therefore perturbations satisfy the equation $\nabla_{\mu} \delta T^{\mu}_{~\nu} = \delta Q_{\nu}$, where $\delta T$ and $\delta Q$ are the perturbations around the background values. To linear order in $\pi$, the $n^\mu$ and $u^\mu$ parts of $\delta Q$ contribute the same and give a term proportional to $\gamma$, as we will see below. Terms differentiating $u$ and $n$ appear to higher order in the clock's equation of motion and generate different three-point function. Nevertheless, to this order they affect the fluid in the same way.
 
  Regarding the fluid, we use a general equation of state parameterized by $w=\overline{p}/\overline{\rho}$ for the background and a sound speed for sound waves in the fluid given by $c_{sr}\equiv\sqrt{\delta p/\delta\rho}$. \footnote{At this point it is good to note that we are considering an effective description of the fluid. Even if in warm inflation for example radiation with $w=1/3$ in the background in appropriate, interaction with the clock or other degrees of freedom could produce a different effective sound speed $c_{sr}^2\neq w$ (for example if $\gamma$ depends on $\rho$) and will also produce a number of terms in the stress-tensor that can be understood using the effective field theory of fluid. That being said, our equation of state here is the one after taking into account all the interactions and therefore is not constrained in principle.}. This equation can be rewritten explicitly in components as 
\bea
&&\delta\dot{\rho}+3 H(1+c_{sr}^2)\delta\rho+a^{-2}(1+w)\rho \partial_i  u_i= -\delta Q_0 \label{ec:u}\\
&&c_{sr}^2 \partial^2 \delta\rho+a^{-3}\partial_{t}[a^3 (1+w)\rho \partial_iu_i]=\partial_i  \delta Q_i\label{ec:u2}
\ea
where the energy flux driven by clock's fluctuations is given to linear order by $\delta Q_i= \gamma N_c \partial_i \pi$ and $\delta Q_0 = 2 \gamma N_c \dot{\pi}$. The quantity $(1+w)\rho\partial_i u^i$ can be eliminated from the expressions above and this gives an equation for the density fluctuations of the fluid as 
\beq\label{eq:radeom}
\delta\ddot{\rho}+[8+3c_{sr}^2]H\delta\dot{\rho}+\left[15  H^2 (1+c_{sr}^2)-c_{sr}^2\frac{ \partial^2}{ a^{2}}\right]\delta\rho=-a^{-2}\partial_i \delta Q_i-\delta\dot{Q}-5H\delta Q,
\eeq
and having the solution of this equation $\delta\rho$, the spatial fluid velocity components $u^i$ can be found from the original equations. The solution can be obtained most easily by going to Fourier space and using $z=-c_{sr}k\eta$ as a time variable
\beq
\frac{d^2\delta\rho}{ dz^2} -\frac{7+3c_{sr}^2}{z} \frac{d\delta\rho}{dz} + \left[\frac{15(1+c_{sr}^2)}{z^2}+1\right]\delta\rho=\frac{1}{z^2H^2}[-a^{-2}\partial_i \delta Q_i-\delta\dot{Q}-5H\delta Q].
\eeq
The solution to this equation has two contributions, the homogeneous solution and the particular one. Since at the linear level the superposition principle holds, the effects of the different sources are decoupled without affecting the results presented here. In the following subsections we will study the homogeneous solution and see how it decays making it irrelevant and we will also study the Green function, which can be derived similarly to the clock's one and dominates the behavior of fluid fluctuations. Finally we will study the longitudinal fluctuations of the frame four-vector $u^\mu$.  

Before going into the details, on the one hand it is clear that for early times space-time is approximately flat and the solution of equation \eqref{eq:radeom} are sound waves with speed $c_{sr}$. On the other hand, after a $\delta\rho$ perturbation exits the horizon, $c_{sr}k_{\rm phys}\rightarrow0$, an additional term proportional to $\delta\rho$ remains in the equation of motion, as opposed to the case of $\pi$. As a result the solution cannot be time independent and it is easy to see that it decays. From the set of equations \eqref{ec:u} and \eqref{ec:u2} its clear that also $\lb u^\mu\rb  \rightarrow \delta^\mu_0$. This is crucial for studying the squeezed limit and will ultimately produce a trivial $f_{\rm NL}$.   

\subsubsection{Homogeneous solution}
~~~~~To understand the dynamics of the fluid we will first solve the continuity equation without any source, for an arbitrary sound speed. This will give us further intuition into the way the fluid responds to sources and also to see how and whether it decays. The equation to solve is
\beq
\delta\ddot{\rho}+[8+3c_{sr}^2]H\delta\dot{\rho}+\left[15  H^2 (1+c_{sr}^2)-c_{sr}^2\frac{ \partial^2}{ a^{2}}\right]\delta\rho=0,
\eeq
which has two independent solutions given by
\beq
\delta\rho_h=A_1z^{5+\nu_r} J_{\nu_r}(z) +A_2z^{5+\nu_r}Y_{\nu_r}(z),
\eeq
where $\nu_r=3c_{sr}^2/2-1$ and $A_1$ and $A_2$ are coefficients fixed by the initial conditions or the initial state of the fluid.

 In the case of radiation with $c_{sr}^2=1/3$, the expression simplifies to $z^{5+\nu_r} J_{\nu_r}(z)\sim z^4\cos{z}$ and $z^{5+\nu_r}Y_{\nu_r}(z)\sim z^4\sin{z}$. This solution shows sound waves with time dependence $$\sin{\left(k\int c_{sr}\frac{dt}{a}\right)},$$ with frequency $\omega=c_{sr}k_{\rm phys}$ and also they redshift with the $a^{-4}$ that characterizes radiation. Since the radiation energy density of the background is constant within the slow-roll approximation (it is constantly being sourced by the clock) the ratio $\delta \rho/\rho$ redshifts as $a^{-4}$, making the fluid fluctuations decay fast. This is in contrast to the case of the radiation driving the background FRW universe, there the background is time dependent and the ratio $\delta \rho/\rho$ evolves differently and may show decaying, constant or growing modes (relative to the background) depending on the details of the problem.
 
  As in the case of the clock, if we put initial conditions at the infinite past $z_0\rightarrow\infty$ then the contribution of this homogeneous solution decay as $z_0^{4}$, due to redshift in the de Sitter background. Then, the accelerated expansion of the universe washes out the information of the initial state of the fluid, just as it does with the clock's initial state. 
  
  Even if the initial conditions were not in the far past, the factor of $z^4$ in the solution will make it decay for $z\ll1$, i.e. when the fluctuations exit the sound horizon $\lambda_{\rm phys} \gg  c_{sr}/H$. This can be seen from the fact that contrary to the clock's case, when $\omega=c_{sr}k_{\rm phys}$ goes to zero, there is still a term proportional to $\delta\rho$ in the equation of motion, impeding a constant non-zero $\delta\rho$ to be a solution.
  
  In the range of $0<c^2_{sr}<1$ the behavior is similar to the radiation, namely sound waves with frequency $\omega= c_{sr} k_{\rm phys}$ and decay due to red-shift in a time-scale of a Hubble. This will make self-interactions of the clock which are mediated by the fluid highly non-local in time. A particular limiting case is non-relativistic matter perturbation for which $c_{sr}^2\simeq0$. Going back to the equation in position space the term with spatial derivatives goes away and the solution at late times is simply 
$
\delta\rho(t,\mathbf{x})\sim a(t)^{-3}f(\mathbf{x}),
$
as expected, where $f(\mathbf{x})$ is the distribution of matter at some initial time. 

\subsubsection{Fluid's Green function}
\begin{figure}[t!]
\begin{center}
\subfigure[$y=100$]{
\includegraphics[scale=0.35]{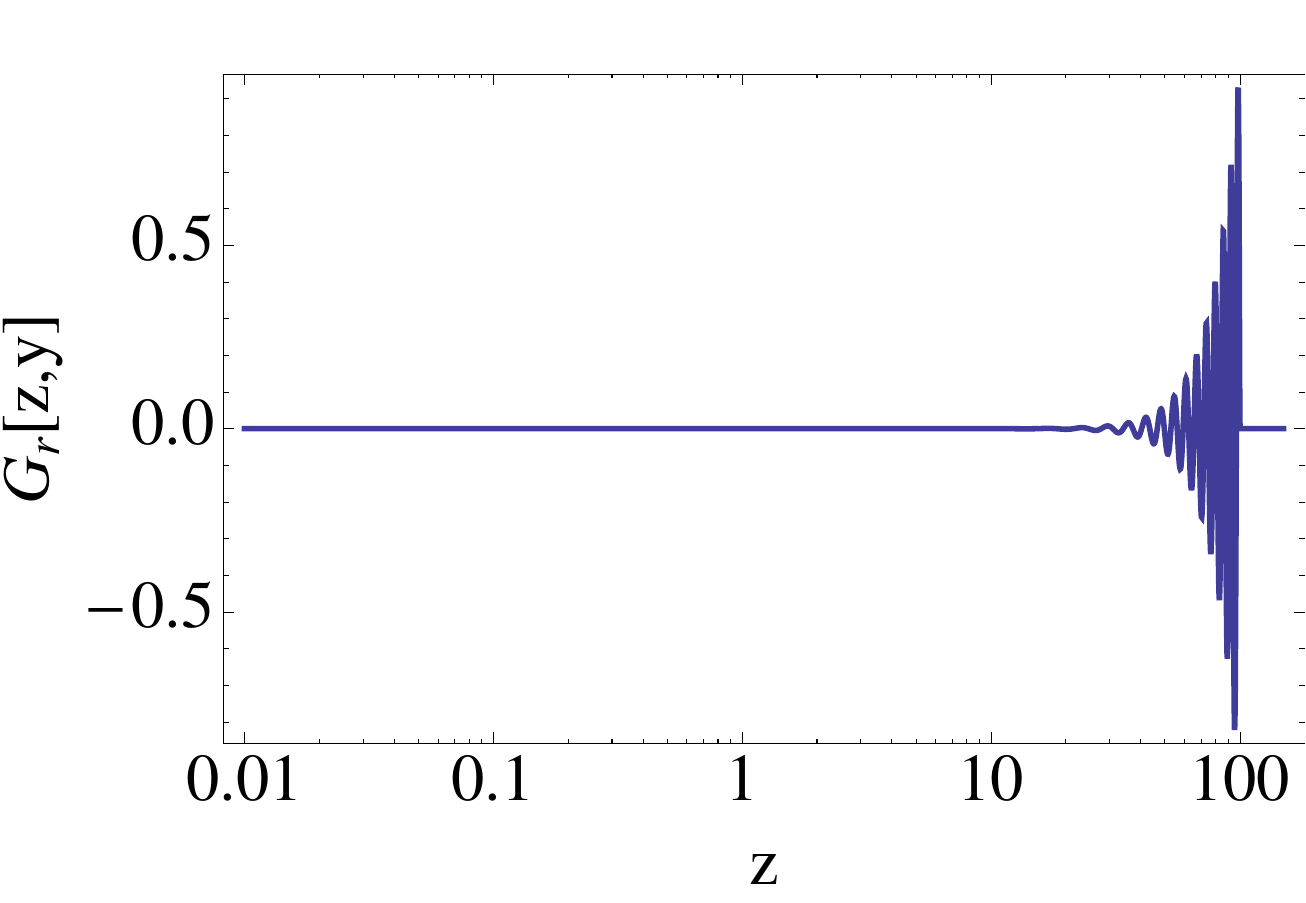}}
\subfigure[$y=10$]{
\includegraphics[scale=0.35]{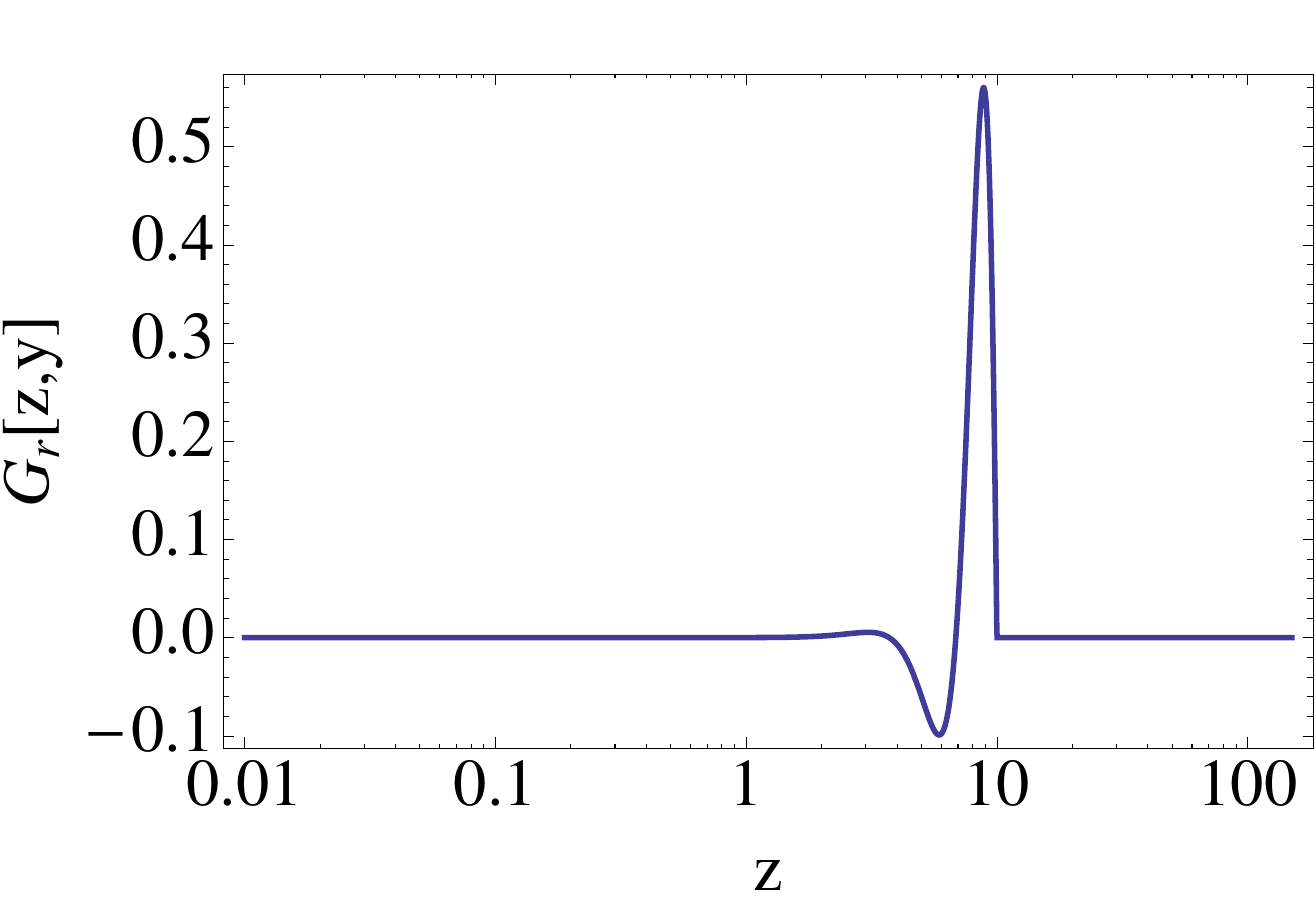}}\\
\subfigure[$y=1$]{
\includegraphics[scale=0.35]{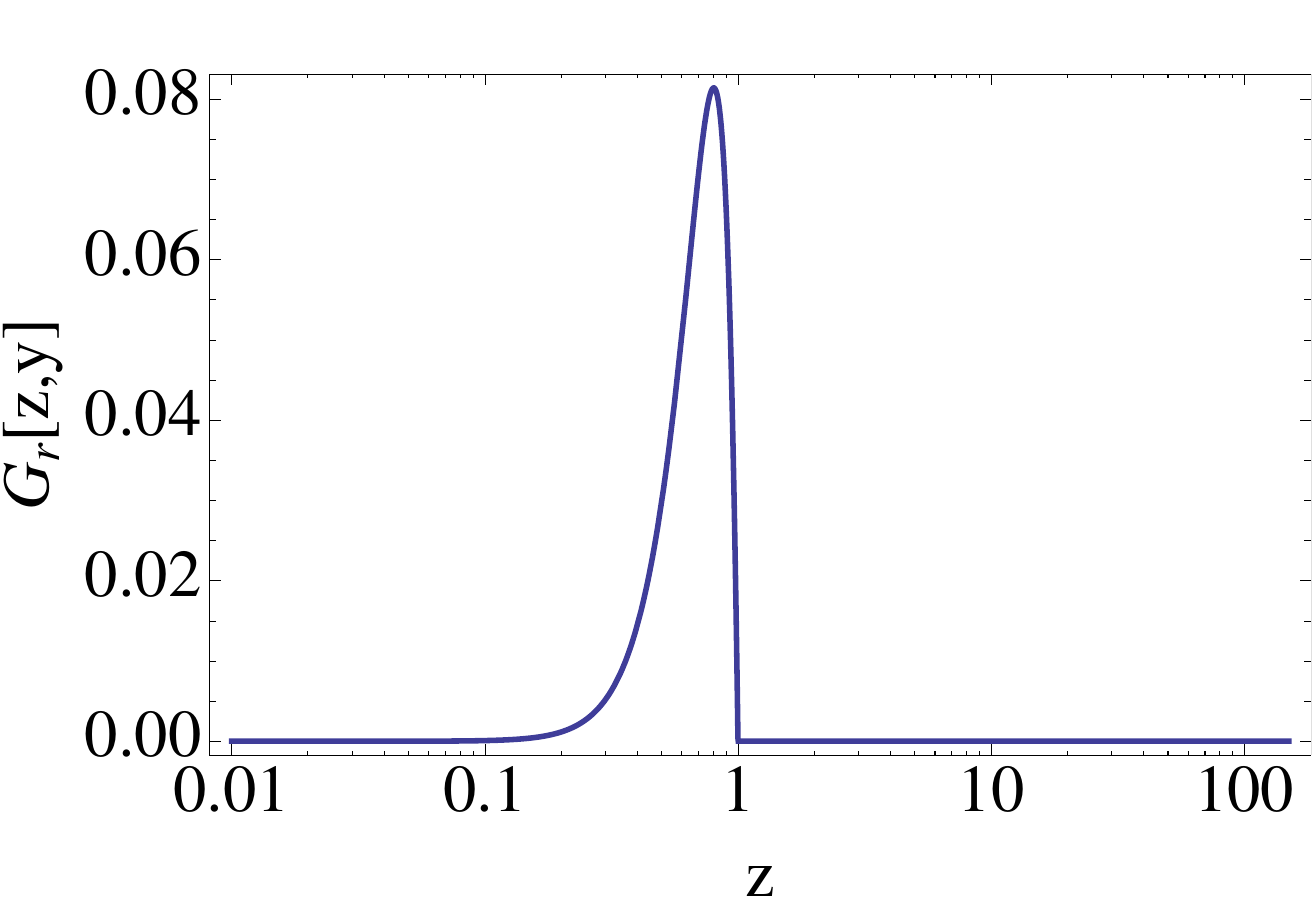}}
\subfigure[$y=0.1$]{
\includegraphics[scale=0.35]{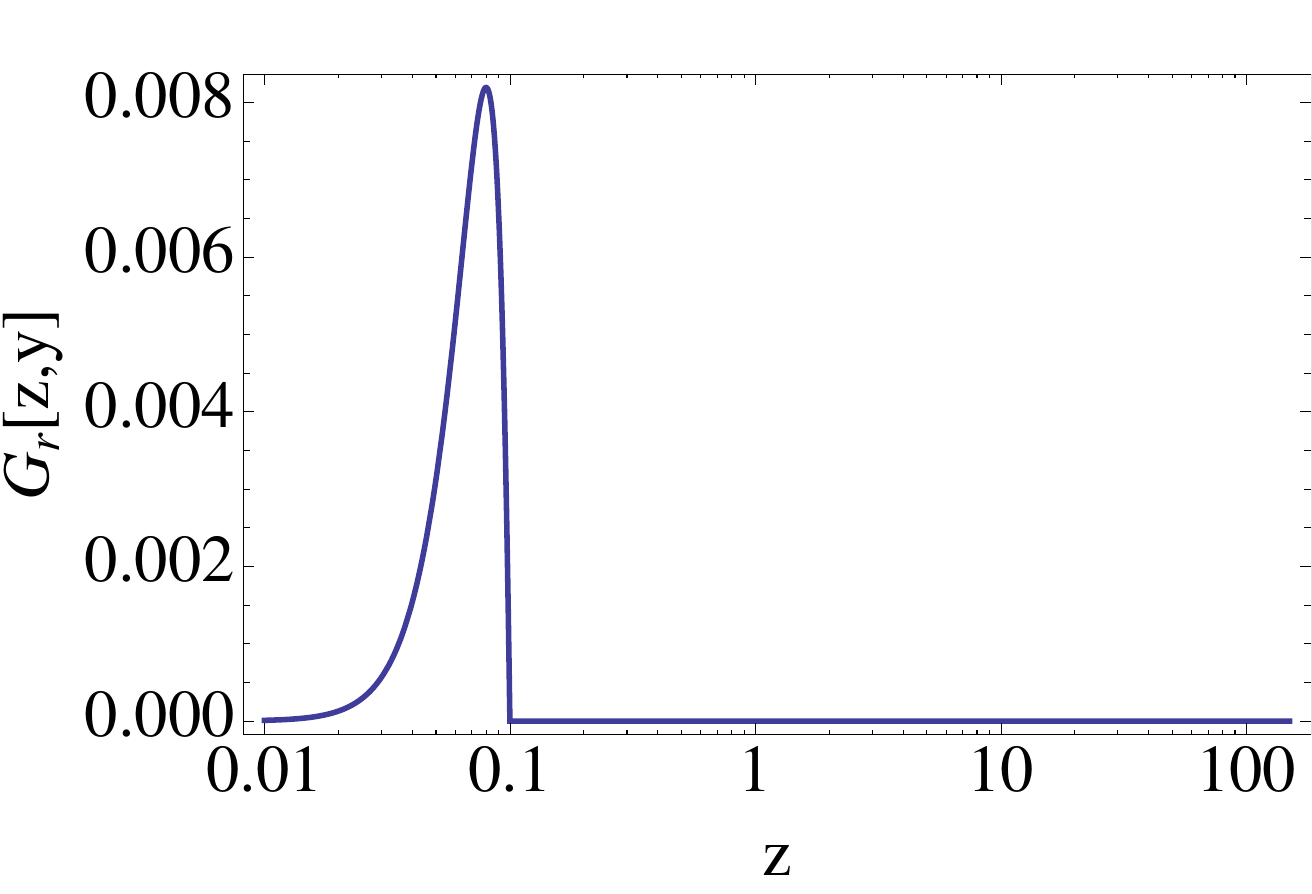}}
\caption{\label{fig:gra} \small\it Plot of the fluid perturbation $\delta\rho$ produced by an instantaneous impulse at four different times $y=-c_{sr}k\eta$, for radiation with sound velocity $c^2_{sr}=1/3$.}
\end{center}
\end{figure}

~~~~~ With the addition of the sources from the clock fluctuations, the solution of the inhomogeneous equation is given by
\beq
\delta\rho = \frac{kc_{sr}}{H^2} \int_{\eta}^{\eta_0}d\eta' G_R(c_{sr}k|\eta|,c_{sr}k|\eta'|)\frac{-a^{-2}\partial_i \delta Q_i-\delta\dot{Q}-5H\delta Q}{(kc_{sr}\eta')^2}\label{eom:rad}
\eeq
with the fluid's Green function 
\beq
G_R(z,z')= z' \frac{\pi}{2} \left(\frac{z}{z'}\right)^{5+\nu_r}[Y_{\nu_r}(z)J_{\nu_r}(z')-J_{\nu_r}(z)Y_{\nu_r}(z')].
\eeq
We will analyze the case of radiation with $c_{sr}^2=1/3$ since it is the one in which the results are more transparent. Using properties of the Bessel functions one can show that the Green function reduces to the result of \cite{moss}, i.e. 
\beq
G_R(z,z')=\left(\frac{z}{z'}\right)^4 \sin{[z-z']}.
\eeq 
This function gives the fluid energy density perturbation at time $t(\eta)$ for a instantaneous source of comoving wavenumber $k$ that acted at time $t'(\eta')<t(\eta)$.This shows a sound wave component $$\sin{\left(k\int_{t'}^{t} c_{sr}\frac{dt}{a}\right)},$$ and also a redshift characteristic of radiation $(a'/a)^4=e^{4H(t'-t)}$. The behavior of this function for sources acting at different times $\eta'$ is shown in figure \ref{fig:gra}. 

One can see how the amplitude of $G(z,z')$ gets redshifted and also that the later the source is turned on, the less significant the effect, especially when the wavelength is outside of the sound horizon. This happens since the period of the sound-wave is $\lambda_{\rm phys}/c_{sr}$. If there were no sound waves, a perturbation would appear and decay following $\sim z^{4}$ independently of the time of generation. Including sound waves, if the perturbation was generated before one period at a time $|\eta'| \lesssim \lambda/c_{sr}$, or equivalently $\lambda_{\rm phys} \gtrsim c_{sr}/H$, then the sound wave would not propagate efficiently. This is due to the fact that in conformal time the period of the wave is fixed and the time interval between the excitation and $t\rightarrow\infty$ is finite. Then the maximum value attainable for the perturbation is $\sim\sin{z'}$, which goes to zero as $z'$ goes to $0$. Equivalently, in cosmic time the period of the sound wave goes to infinity. This attenuation is enhanced by the redshift of the perturbation. One can interpret this realizing that a source which is homogeneous over the sound horizon cannot affect the fluid; all of this follows directly from space-time locality.

Having characterized the response of the fluid through its Green function now we analyze the sources of sound-waves. We will write them in terms of the clock fluctuation $\pi$, the equation of motion for $\delta\equiv\delta\rho/3(\rho+p)$ and neglecting slow-roll suppressed terms. We get: 
 \beq\label{eq:radt}
\ddot{\delta}+[8+3c_{sr}^2]H\dot{\delta}+\left[15  H^2 (1+c_{sr}^2)-c_{sr}^2\frac{\partial^2}{a^2} \right]\delta=H\left[ -\frac{\partial^2}{a^2} \pi -2\ddot{\pi}-10H\dot{\pi}\right],
\eeq
 which has the solution in comoving Fourier space given by
 \beq\label{deltaT}
\delta=\frac{kc_{sr}}{H}\int_{\eta}^{\eta_0}d\eta' G_R(c_{sr}k|\eta|,c_{sr}k|\eta'|)\frac{\left[\frac{k^2}{a^2}\pi-2\ddot{\pi}-10H\dot{\pi}\right]}{(c_{sr}k\eta')^2}.
 \eeq
 
Therefore the sources for fluid fluctuations, as seen from equation \eqref{eq:radeom} and \eqref{eq:radt}, are $\partial^2\pi, \ddot{\pi}$ and $\dot{\pi}$. In the next section we will see how non-Gaussianities and in particular the three-point function are driven by correlations of the type $\lb \partial^2 \pi \pi_f\rb$, $\lb \ddot{\pi} \pi_f \rb$ and $\lb \dot{\pi} \pi_f\rb$ through the presence of correlations such as $\lb \delta \rho_{\mathbf{k}}\pi_{\mathbf{k}f}\rb$ and $\lb u^i_{\mathbf{k}} \pi_{\mathbf{k}f} \rb$. This has several implications. First of all, fluid fluctuations are going to be dominated by spatial inhomogeneities of the clock since the correlations of $\pi_f$ with $\dot{\pi}$ and $\ddot{\pi}$ are strongly suppressed with respect to $\lb \partial^2 \pi \pi_f\rb$. As we explained in the last section this source starts acting at freeze-out with decreasing amplitude until the mode exits the Hubble horizon. On the other hand, starting from the freeze-out time of the mode, the fluid will start responding with red-shifting sound-waves. When the mode exits the sound horizon (which happens before the mode exits the Hubble horizon) the fluid fluctuation will start decreasing without showing sound-waves and at late times $\delta \rho \rightarrow 0$. As explained before this is due to the finite time interval of conformal time, which after sound horizon crossing it is smaller than the wave period. 

Even if the energy density $\delta T^{00}$ is driven by $\dot{\pi}$, the source $\partial_i \pi$ is relevant because fluctuations of $u^i$ and $\delta \rho$ are coupled. From an EFT point of view, if one has a non-local term sourced by $\partial^2\pi$ for example, the realization of coordinate invariance of the ADOF forces source terms like $\dot{\pi}$ and $\ddot{\pi}$ to appear in a specific combination.     

\subsubsection{Velocity Perturbation}
~~~~~ Having the solution for the density perturbation one can obtain the linear order velocity field of the fluid's frame $u^\mu$. If the diffeomorphism invariance of the friction term in the equations of motion is realized in the form $\partial_t\rightarrow u^\mu\partial_\mu$, understanding how this quantity behaves is important to understand the effects of the non-linearities it generates. We will compute the contribution from the longitudinal sound waves of the fluid (in principle, there could be vectorial contributions like vortices but we will not consider them here, mainly because they get red-shifted away more quickly than longitudinal modes). For that purpose we will decompose $u^i$ as 
\beq
u_i=\partial_i u_L + {u_i}_V
\eeq
Using the fluid equation of motion we can find and motivate a non-local behavior that characterizes sound waves. Working as before in Fourier space for comoving coordinates, plugging equation \eqref{deltaT} in equation \eqref{ec:u} gives  
\bea
a^{-2}\partial_i u_i&=&a^{-2}\partial^2 u_L =-3( \dot{\delta} + 3H(1+c_{sr}^2)\delta)+6H\dot{\pi}\nn
&=& -3k c_{sr} \int_\eta^{\eta_0} d\eta' (c_{sr}k|\eta|)G_V(c_{sr}k\eta,c_{sr}k\eta') \frac{\left[\frac{k^2}{a^2}\pi-2\ddot{\pi}-10H\dot{\pi}\right]}{(c_{sr}k\eta')^2}+6H\dot{\pi}
\ea
and then the scalar velocity $u_L$ can be written as 
\beq\label{eq:ul}
u_L=c_{sr}^2 \frac{3k c_{sr}}{H^2} \int_\eta^{\eta_0} d\eta' \frac{G_V(c_{sr}k\eta,c_{sr}k\eta')}{c_{sr}k|\eta|} \frac{\left[\frac{k^2}{a^2}\pi-2\ddot{\pi}-10H\dot{\pi}\right]}{(c_{sr}k\eta')^2}-\frac{6}{(k\eta)^2}\frac{\dot{\pi}}{H}.
\eeq
Taking the spatial derivative one can find the longitudinal component of velocity fluctuations as $u_i = -i k_i u_L$. The physical velocity of the fluid elements can be obtained as 
\beq\label{eq:u}
u^i_{\rm phys}=au^i =-i\hat{k}_i \frac{3k c^2_{sr}}{H} \int_\eta^{\eta_0} d\eta'G_V(c_{sr}k\eta,c_{sr}k\eta') \frac{\left[\frac{k^2}{a^2}\pi-2\ddot{\pi}-10H\dot{\pi}\right]}{(c_{sr}k\eta')^2}+ik_i\frac{6}{k^2}H\dot{\pi}.
\eeq
In all the expressions above, the velocity's Green function is given by 
\bea
G_V(c_{sr}k\eta,c_{sr}k\eta') &=&\left[ \partial_z - 3(1+c_{sr}^2)\frac{1}{z}\right]G_r(c_{sr}k\eta,c_{sr}k\eta')\nn
&=&\frac{\pi }{2} y \left(\frac{z}{y}\right)^{5+\nu_r} \left[J_{\nu_r+1}(z)Y_{\nu_r}(y)-J_{\nu_r}(y)Y_{\nu_r+1}(z)\right]\nn
&\rightarrow&\left(\frac{z}{y} \right)^4\cos{[z-y]}~~~~\text{for radiation}
\ea 
\begin{figure}[t!]
\begin{center}
\subfigure[]{
\includegraphics[scale=0.6]{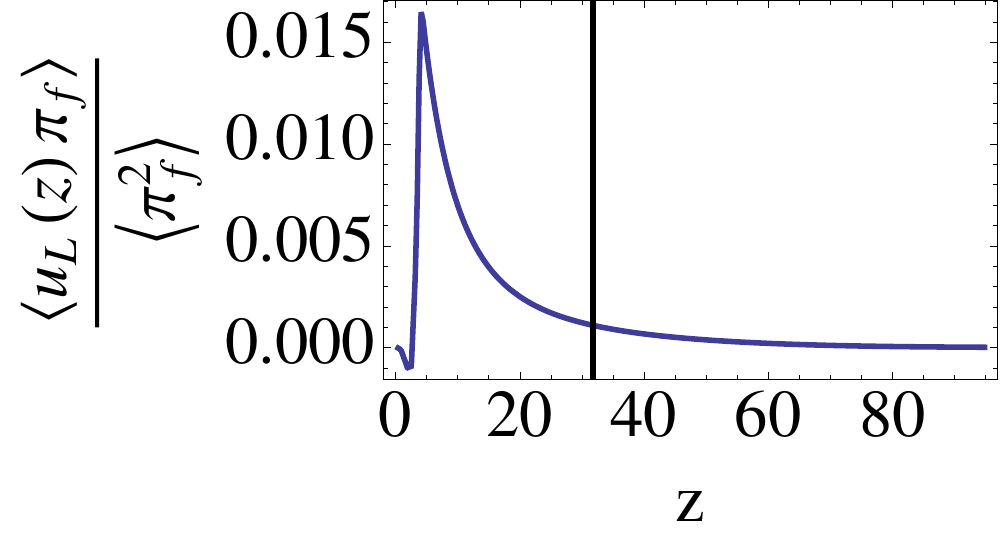}}\\
\subfigure[]{
\includegraphics[scale=0.57]{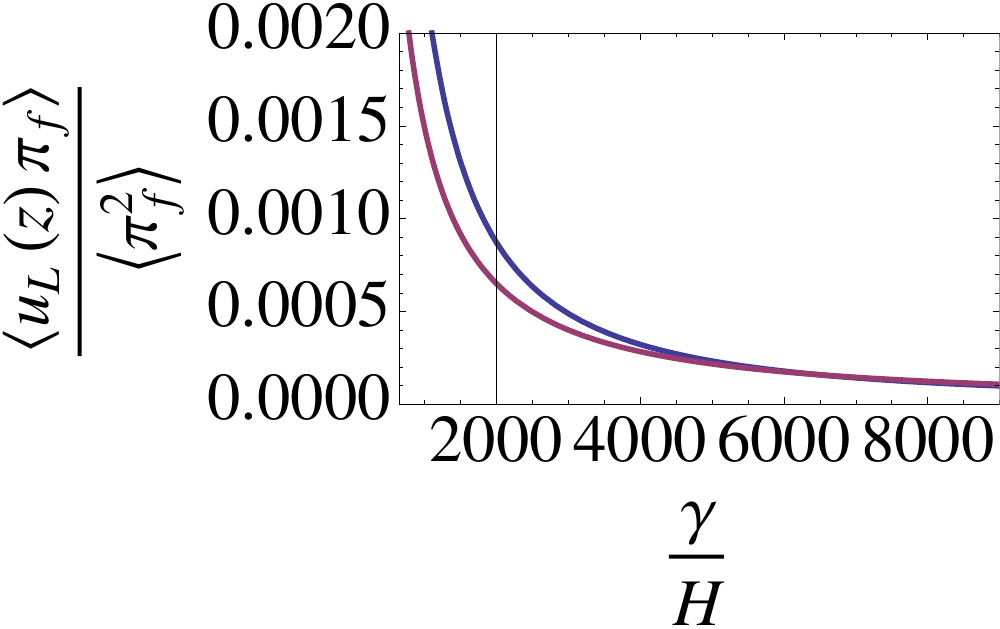}}
\caption{\label{fig:upi} \small\it (a) Plot of the correlation $\lb u_L(z)\pi_f\rb$ as a function of $z$ for $\gamma=10^3H$. The freeze-out time is located at $z=z_\star\sim 30$. (b) Plot of the correlation $\lb u_L(z)\pi_f\rb$ evaluated at freeze-out as a function of $\gamma/H$ (blue curve). Approximate curve decaying as $(\gamma/H)^{-1}$ (red curve).}
\end{center}
\end{figure}
The Green function that determines the non-local part has the same properties as the one for the density perturbations regarding the sound waves sourced by $\pi$. The only difference is that they are $90^\circ$ out of phase with density fluctuations and that velocity perturbations redshift faster or slower depending if one considers $u_L$, $u_i$ or $u^i$ but the physical speed redshifts with the $a^{-4}$ dependence. We will see in the next section that non-Gaussianities will be driven by correlations like $\lb u_{\mathbf{k}L} \pi_{\mathbf{k}f}\rb$. From the expression \eqref{eq:ul} it is clear that when one goes well inside the horizon there is an extra factor of $k_{\rm phys}$ in the Green function that suppresses this correlation with respect to, for example $\lb \pi^2 \rb$. This will compete with the enhancement factor from the noise-induced non-Gaussianity and it will give a reduce $|f_{\rm NL}|$ in the strong dissipation regime, with respect to the contribution from local dissipation.

 To see this explicitly, figure \ref{fig:upi} shows the behavior of $\lb u_{\mathbf{k}L}(z) \pi_{\mathbf{k}f}\rb$ (as a function of $z$) and $\lb u_{\mathbf{k}L}(z_\star) \pi_{\mathbf{k}f}\rb$ (as a function of $\gamma$). First of all, the correlation peaks at horizon exit. This is because the source $\lb \partial^2 \pi \pi\rb$ acts from freeze-out to horizon exit. After horizon exit the correlation decays because the source strength decays. The value of the correlation at freeze-out will be relevant for the equilateral $f_{\rm NL}$ and from the second plot one can see that it decays as $\sim k^{-2}_{\rm phys} \sim (\gamma/H)^{-1}$. 

There is also a local part given by the last term on the right hand side of equation \eqref{eq:u} that as we will see would lead to a non-linear term on the clock equation of motion of the form $$u^i_{\rm local}\partial_i\pi\sim\nabla^{-1}\partial_i\dot{\pi} \partial_i \pi.$$ Again, this combination of sources is fixed by the symmetries of the fluid dynamic. Nevertheless, because of the behavior of $\pi$ correlations explained in the last section this contribution is sub-leading since it is driven by time derivatives and also is suppressed by a $k_{\rm phys}^{-2}\sim \gamma^{-1}$ factor making it negligible. To next-to-leading order in slow-roll the long wavelength mode can be approximated by $\dot{\pi}\sim \epsilon H \pi$ and this term could give a non-vanishing contribution to the squeezed limit, without affecting the tilt of the two-point function. This could affect the consistency condition. Nevertheless, one can integrate by parts the non-local term with two time derivatives in \eqref{eq:ul} and this local term cancels.

\subsection{Perturbations deep inside the horizon}\label{sec:flat}
~~~~~Before moving to the full computation of the three-point function we will compute it in the approximation of flat space. This is a useful estimate because for strong dissipation $\gamma\gg H$, and at the time of freeze-out during which the relevant interaction is happening $\lambda_{\rm phys}\ll H^{-1}$. 

For interactions local in time (like those of \cite{diseft}) this approximation becomes exact in the strong dissipation regime. In the case studied in this paper the interaction is non-local on the scale of a Hubble time that set the timescale for the redshift of the fluid. This will change the flat space result by an amount that does not disappear in the $\gamma\gg H$ limit. Nevertheless, a ``zeroth order" effect of the sound waves on the three-point function can be understood in terms of the flat space result. An example is the suppression of $f_{\rm NL}$ with respect to the local case.

Regarding the approximate expressions for the clock's perturbation we can cite the results from \cite{diseft}, which are valid in the strong dissipation regime. Mainly, the expression for $\pi$ is given by
\beq
\pi_k(t)=-\frac{1}{N_c}\int_0^\infty dt' G_\gamma^k(t-t')\delta\calo_{\mathcal{S}}(t'),
\eeq
where 
\beq
G_\gamma^k(t-t')=\frac{1}{\gamma}e^{-\frac{\omega_0^2}{\gamma}(t-t')}\Theta(t-t'),
\eeq
and we define $\omega_0=c_s k=c_s k_{\rm comov.}/a$ which in this approximation we take to be a constant. The two point function of the stochastic source is given by 
\beq
\lb \delta\calo(\mathbf{k},t)\delta\calo(\mathbf{q},t')\rb=\nu_\calo \delta(t'-t)(2\pi)^3\delta(\mathbf{k}+\mathbf{q}).
\eeq

Regarding the fluid, the equation for $\delta(\mathbf{k},t)$ in flat space is given by 
\beq
\left(\partial_t^2-c_{sr}^2\partial^2\right)\delta\simeq -\partial^2 \zeta,
\eeq
and the solution is 
\beq
\delta_k(t)= \int_0^\infty G_R^k(t-t')k^2\zeta(t'),~~~~\text{with}~~G_R^k(t-t')=\frac{\sin{c_{sr}k (t-t')}}{c_{sr}k}\Theta(t-t').
\eeq
Using the same procedure as we did before, we can find the scalar velocity perturbation $u^i=\partial_i u_L$ as 
\beq
k^2 u_L = 3\dot{\delta}_k(t),~~~\text{then}~~~u_L=\int_0^\infty G_V^k(t-t')k^2\zeta_k(t'),
\eeq
where the Green function is given by 
\beq
G_V^k(t-t')=\frac{3\cos{c_sk(t-t')}}{k^2}\Theta(t-t').
\eeq

Using these approximate expressions we can now compute the non-Gaussian three-point function. We will do it in the equilateral limit for two reasons. First the three-point function is peaked there and second, when $k_1\simeq k_2 \simeq k_3 \equiv k$ the interaction takes place when all of them are freezing-out and the flat space approximation of this section is valid for the three modes. 

As we will see in the next section the most important contribution from the fluid comes from a non-linear term in the equation of motion of the form $\gamma_u u^i \partial_i \pi$. Therefore the three-point function is  
\bea
\lb \zeta_k \zeta_k\zeta_k\rb&=&6 \frac{\gamma_u}{H}  \int dt' G_\gamma^k(t-t')k^2 \lb u_L(t') \zeta_k\rb \lb \zeta_k(t')\zeta_k\rb\nn
&=&6 \frac{\gamma_u}{H} \frac{\nu_\calo^2H^4}{N_c^4} \int dt' G_\gamma^k(t-t')k^2 \int dt'' G_V^k(t'-t'')k^2\nn
&&~~\times \int dt'''' G_\gamma^k(t''-t'''')G_\gamma^k(t-t'''')  \int dt''' G_\gamma^k(t'-t''')G_\gamma^k(t-t''')
\ea
For reference for the next section we will compute in this limit the parameter 
\beq
f_{\rm NL}=\frac{5}{6}\frac{\lb \zeta_k \zeta_k\zeta_k\rb}{3 \lb \zeta_k\zeta_k\rb},
\eeq
which is given by
\bea
|f_{\rm NL}|(t)&=&\frac{5}{3}(\omega^2\gamma)^2 \frac{\gamma_u}{H}  \int dt' G_\gamma^k(t-t')k^2 \int dt'' G_V^k(t'-t'')k^2\nn
&&~\times\int dt'''' G_\gamma^k(t''-t'''')G_\gamma^k(t-t'''')  \int dt''' G_\gamma^k(t'-t''')G_\gamma^k(t-t''')\nn
&&\rightarrow 5\frac{\gamma^2}{c_s^2(c_sk)^2+c_{sr}^2\gamma^2}
\ea
Evaluated at freeze-out $c_s k_\star \simeq \sqrt{\gamma H}$ this estimation gives the result 
\beq
|f_{\rm NL}|\sim \frac{\gamma_u}{Hc_s^2}~\frac{5}{1+\frac{c_{sr}^2}{c_s^2}\frac{\gamma}{H}}
\eeq
\begin{figure}[t!]
\begin{center}
\includegraphics[scale=0.7]{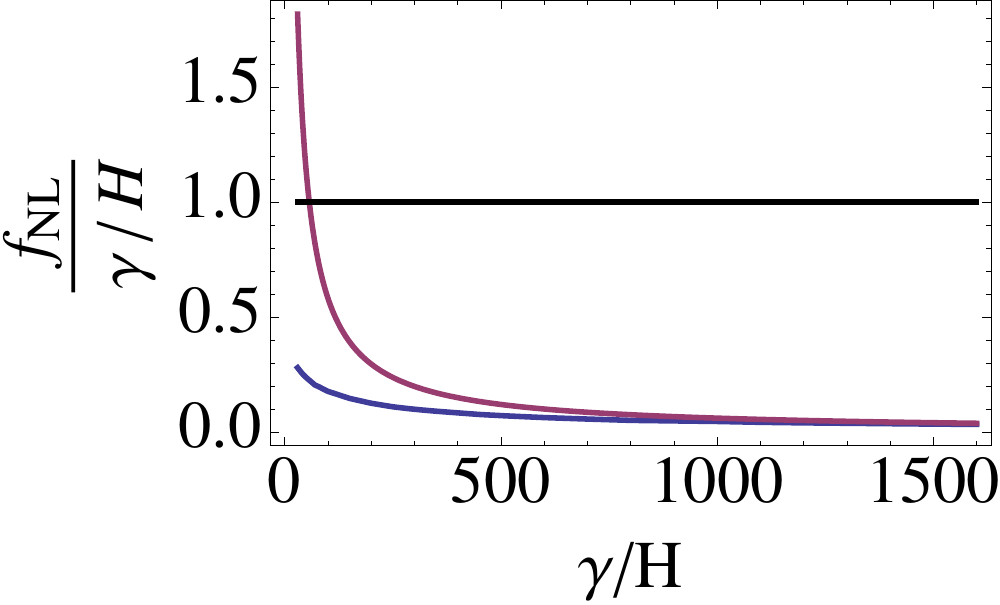}
\caption{\label{fig:compflat} \small\it  Plot of $f_{\rm NL}H/\gamma$ for local dissipation (black curve), fluid in flat space (red curve) and for completeness we also show the complete result for the fluid in deSitter (blue curve).}
\end{center}
\end{figure}

On the other hand, the source for the three-point function when there is only one preferred frame with local dissipation comes from a term in the equation of motion of the form $$\gamma_n n^i\partial_i \pi=\gamma_n\partial_i \pi \partial_i \pi,$$which was studied in \cite{diseft}. Computing the parameter $f_{\rm NL}$ for this term in the same approximations gives 
\bea
|f_{\rm NL}|(t)&=&\frac{5}{3}(\omega^2\gamma)^2 \frac{\gamma_n}{H}  \int dt' G_\gamma^k(t-t')\left(\int dt'' G_\gamma^k(t'-t'')G_\gamma^k(t-t'') \right)^2\nn
&&\rightarrow \frac{\gamma_n}{Hc_s^2}.
\ea

Then, in the flat space approximation the conclusion is that $f_{\rm NL}$ grows linearly with $\gamma$ for the $n^\mu$ term while the $u^\mu$ term grows but more slowly. This can be seen in figure \ref{fig:compflat} where we also show the full result in de-Sitter with the fluid for completeness. For the interaction $n^\mu \partial_\mu \pi$ the flat space result reproduce completely the full three-point function as long as $\gamma/H\sim (k/aH)^2 \gg 1$. 

We will see in the next section that the full computation is indeed slightly different from the flat space result, but the latter gives a good estimate for the order of magnitude 
\beq
f_{\rm NL} \sim \frac{\gamma_u}{\gamma}\frac{1}{c_{sr}^2},
\eeq
for reasonable values of $\gamma/H$. Clearly, the fact that non-local dissipation (or equivalently with more than one preferred frame) produces non-Gaussianities which are suppressed with respect to the local dissipation one (or with only one preferred frame) can be understood without including curvature. 

\section{Non-Gaussianities}\label{sec:ng}
~~~~~In this section we will go into the details of the calculation of non-Gaussianities. As discussed in \cite{diseft}, the possible non-linear terms that can appear in the equation of motion for the clock  are not as constrained as the form of the two-point function and this freedom allows for a variety of behaviors to happen. We will analyze the shape of the three-point function, checking along the way that the squeezed limit is trivial in accordance with the consistency condition of the three-point function, since we are working to zeroth order in slow-roll. This will make the arguments in \cite{diseftcon} applicable without any change and then the consistency condition will be satisfied.  

\subsection{Non-Gaussianities from the response}
~~~~~We will first study the non-Gaussianities generated by non-linear terms in the response which appear to realize diffeomorphism invariance. Since terms coming from $\partial_t\rightarrow n^\mu\partial_\mu$ have been studied in \cite{diseft} we will just quote the result of that work. The shape of the three-point function has a peak on the equilateral configuration and a smaller peak around $x_2\simeq x_3\simeq1/2$ of opposite sign. The strength is characterized by the value corresponding to the equilateral triangle and it is given by $|f_{\rm NL}^{\rm eq}|\simeq \gamma_n/4Hc_s^2$. In this work we will study non-linear terms coming from $\partial_t\rightarrow u^\mu\partial_\mu$, which generate a non-local response mediated by sound waves in the ADOF and give a new contribution to the three-point function. 
\subsubsection{Non-linear covariance realization}
~~~~~The term $\gamma_u u^\mu\partial_\mu \pi$ in the equation of motion generates non-linear terms of several types. The leading order perturbation comes from the term $\gamma_u u^i \partial_i \pi$, with $u^i=u_{(1)}^i+u_{(2)}^i+\ldots$, where the subscript indicates the order in $\pi$ of the response and for the background $u_{(0)}^i=0$. The main contribution will be given by the first order term corresponding to linear response, since the second or higher order terms contributing to $u^i$ either do not contribute to the three-point function or are supressed. There are also terms coming from $\gamma_u \delta u^0 \partial_t \pi$ but the leading contribution to $\delta u^0$ is second order in the $u^i$ response, since $u^2=-1$, and therefore second order in $\pi$ giving a term in the equation of motion schematically of the form $\sim \gamma_u \pi^2 \dot{\pi}$ which does not contribute to the three-point function. There are also terms coming from metric perturbations which are slow-roll suppressed and thus are also sub-leading.

Taking this into account the main contribution to non-Gaussianities will be from a $\gamma_u u_{(1)}^i \partial_i \pi$ interaction which we will study in detail below (from now on we will omit the subscript since we will deal always with linear response). Using the expression for $u[\pi]$ gives a non-linear term in the equation of motion of the form
\beq
\gamma_u u^i[\pi] \partial_i \pi\sim \gamma_u \int G_R [\{\partial^2, \partial_t,\partial_{tt} \}\partial_i\pi \partial_i \pi]+\gamma_u \partial_i\dot{\pi}\partial_i \pi.
\eeq 
There are four different kinds of terms contributing to this non-Gaussianity, three non-local sourced by $\partial^2\pi, \dot{\pi}$ and $\ddot{\pi}$ and a local one sourced by $\partial_i \dot{\pi}$. In the strong dissipation regime the terms involving time derivatives will be suppressed by $\gamma/H$ factors and are therefore sub-leading. This was explained in \cite{diseft} and it is due to the fact that time derivatives scale as powers of $H$ rather than $\sqrt{\gamma H}$. Therefore, the only relevant term left is the perturbation in the fluid sourced by spatial derivatives $\partial^2\pi$ and we will concentrate on this leading contribution. 

To compute the strength of the non-Gaussian three-point function we need the leading order contribution to the fluctuations of the clock given by the non-linear term in the equation of motion. Using linear response it is given by 
\beq
\pi_2(k,\eta\rightarrow0)=\frac{\gamma_u}{H^2} \int d(kc_s\eta') G_\gamma(0,k c_s |\eta'|)\frac{[u^i \ast \partial_i \pi](\mathbf{k},\eta')}{(kc_s\eta')^2} ,
\eeq
where the asterisk means taking the convolution with wave-vector $\mathbf{k}$. Using this expression one can compute the three-point function as the average of the product of clock fluctuations as
\bea
\lb \zeta (\mathbf{k}_1) \zeta (\mathbf{k}_2) \zeta (\mathbf{k}_3) \rb &=& \sum_{\rm cyclic} H^3 \lb \pi_2(\mathbf{k}_1) \pi_1(\mathbf{k}_2) \pi_1(\mathbf{k}_3)\rb\nn
&\equiv& (2\pi)^3 \delta^{(3)}(\mathbf{k}_1+\mathbf{k}_2+\mathbf{k}_3)F(k_1,k_2,k_3),
\ea
and we defined the function $F(k_1,k_2,k_3)$ as usual \cite{shape}. Using the response for the fluid velocity gives the following expression in term of the known two-point functions 
\beq\label{eq:vtermng}
F(k_1,k_2,k_3)=\sum_{\rm perm.}\frac{\gamma_u}{H^2}  \int d(k_1c_s\eta') \frac{G_\gamma(0,k_1 c_s |\eta'|)}{(k_1c_s\eta')^2}\lb u^i(k_2,\eta') \zeta_f(k_2)\rb \lb\partial_i \zeta(k_3,\eta')\zeta_f(k_3)\rb
\eeq
In appendix \ref{app:ng} we give an explicit expression in terms of the Green functions of the fluid and clock, which we used to perform a numerical calculation.

 Following the standard analysis of \cite{shape} we set an arbitrary scale $k$, define $x_i=k_i/k$, compute $F(x_1,x_2,x_3)=k^6F(x_1k,x_2k,x_3k)$ and then chose $k=k_1$ making $x_1=1$. Therefore, the information of the shape of the non-Gaussianity is encoded in the function $F(x_2,x_3)=x_2^2x_3^2F(1,x_2,x_3)$ for $x_3\leq x_2$ and $x_3\geq1-x_2$. 
 \begin{figure}[h!]
\begin{center}
\subfigure[]{
\includegraphics[scale=0.5]{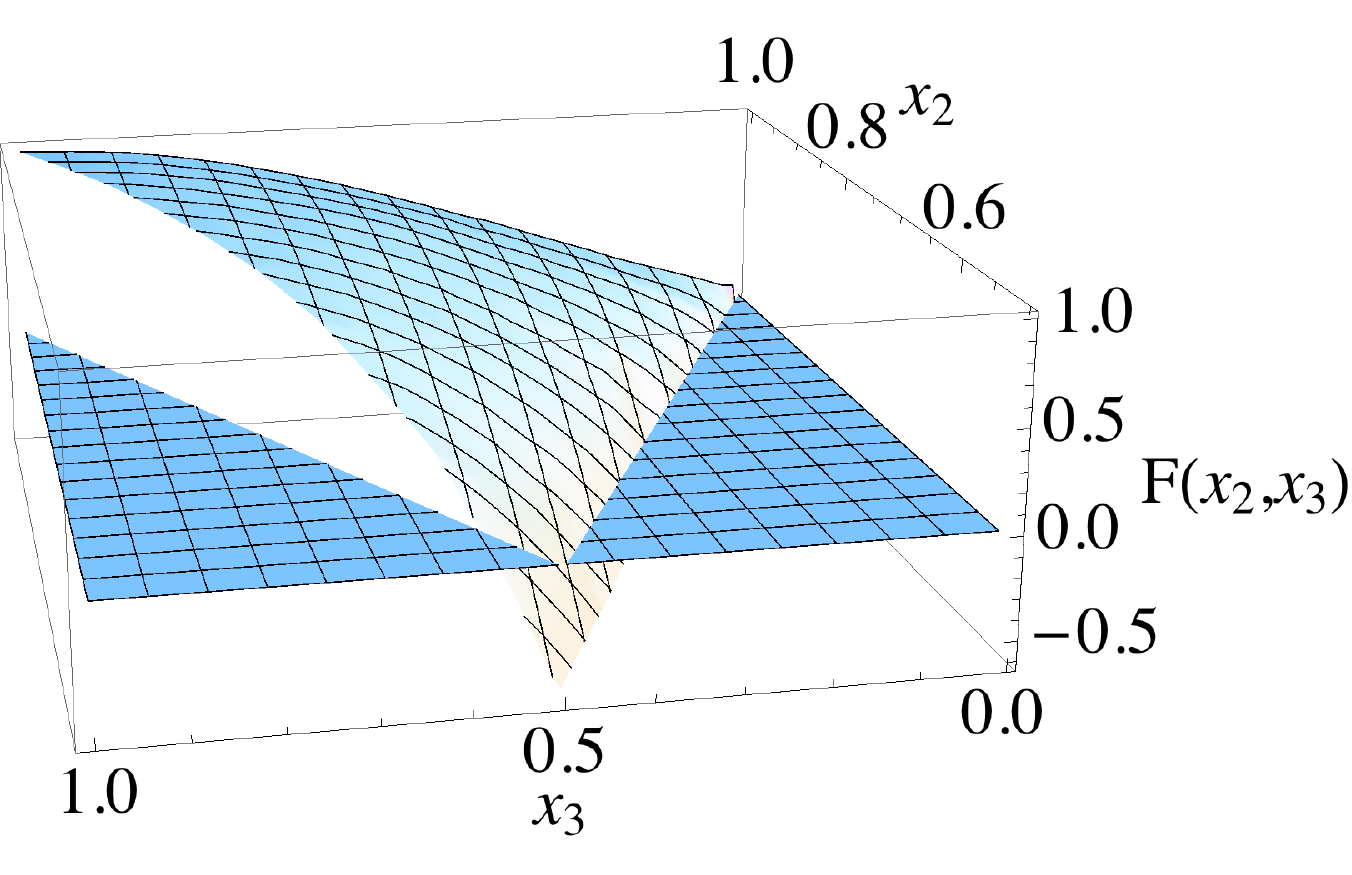}}
\subfigure[]{
\includegraphics[scale=0.6]{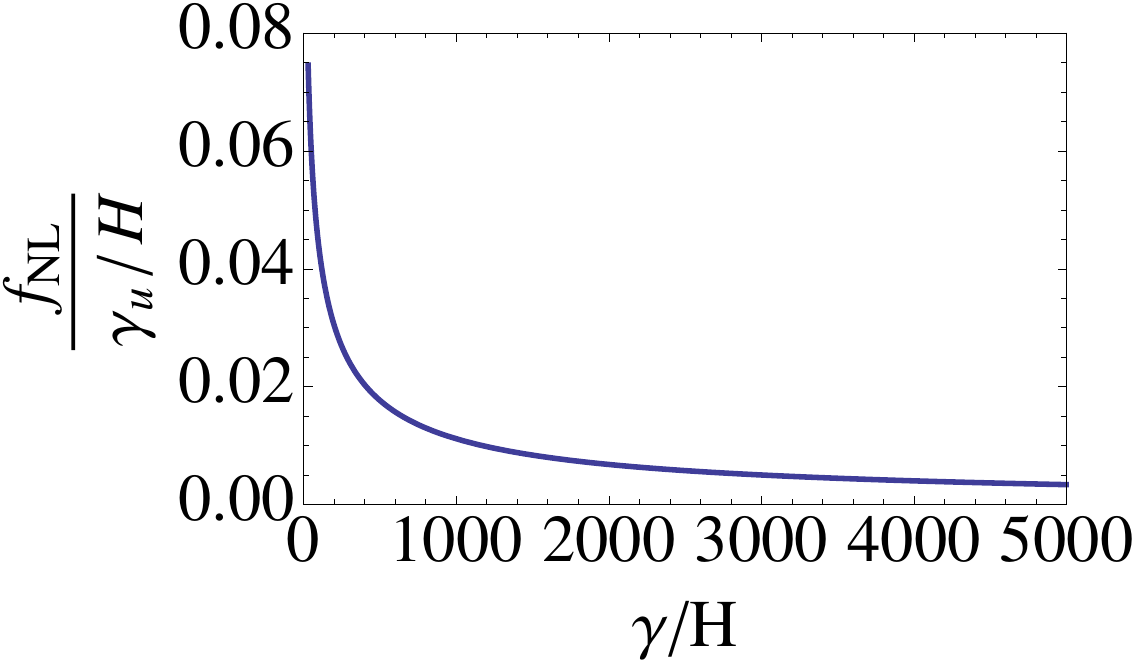}}
\subfigure[]{
\includegraphics[scale=0.5]{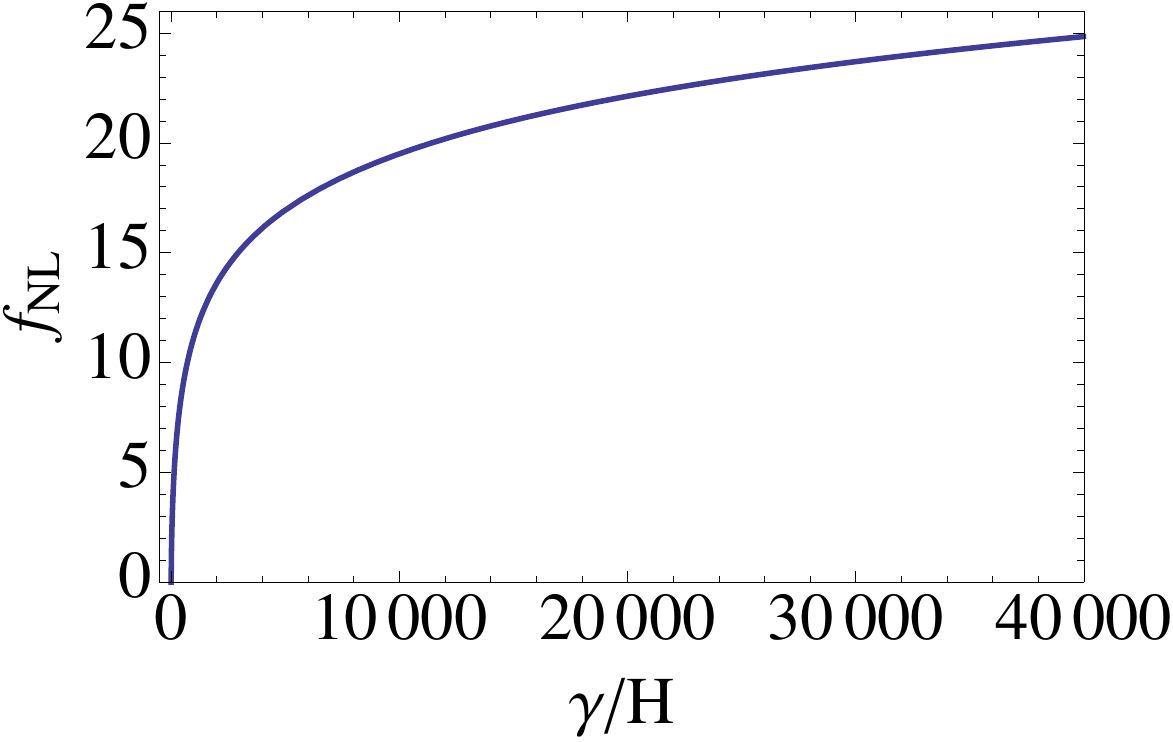}}
\caption{\label{fig:Fv1} \small\it (a) Plot of the non-Gaussianity $F(x_2,x_3)=x_2^2x_3^2F(1,x_2,x_3)/F(1,1,1)$ computed numerically with the exact expression for the term $\gamma_u u^\mu\partial_\mu\pi$ for a dissipation of $\gamma/H=10$ and a relativistic fluid. The shape for different dissipation is similar. (b) Plot of the strength of the non-Gaussianity $f_{\rm NL}/(-\gamma_u/H)$. (c) Plot of the parameter $|f_{\rm NL}|$ for the case $\gamma_u=\gamma$, $\gamma_n=0$.}
\end{center}
\end{figure}
The plot of the shape of this non-Gaussianity is shown in figure \ref{fig:Fv1}(a). The overall shape is very similar to the one found for the $\gamma_n n^\mu\partial_\mu\pi$ term. Following \cite{shape} the way to characterize a given non-Gaussian shape is to give its cosine with fiducial shapes such as equilateral $F_{\rm equil}(x_2,x_3)$, orthogonal $F_{\rm ortho}(x_2,x_3)$ and local $F_{\rm local}(x_2,x_3)$. For the interaction studied in this section it was found that the cosines are 
\bea
&&\cos\theta_{\rm equil}\simeq0.95,\\
&&\cos\theta_{\rm ortho}\simeq0.49, \\
&&\cos\theta_{\rm local}\simeq0.37.
\ea
These cosines were computed for $\gamma=40H$ and $c_{sr}=\sqrt{1/3}$ although the shape is nearly independent of these parameters. Instead of the cosines one could give the ``fudge factors" writing the shape in the following way
\beq
\frac{F(x_1,x_2,x_3)}{F(1,1,1)}\simeq0.41F_{\rm equil}+0.16F_{\rm ortho}+0.04F_{\rm local}+\ldots,
\eeq
where the dots represent other possible fiducial shapes not considered here. An approximate template for this shape is given by 
\bea
F(x_1,x_2,x_3)&=&\frac{P^2_\zeta f_{\rm NL} }{32 x_1^3 x_2^3 x_3^3} \bigg[\frac{1532}{99} (x_1^3-x_2^3-x_3^3)-\frac{1532}{99} \left(x_1^3+x_2^3-x_3^3\right)\nn
&&-\frac{1532}{99} \left(x_1^3-x_2^3+x_3^3\right)+4\frac{ x_1^2 \left(x_2^2+x_3^2\right)+x_2^2 x_3^2}{x_1+x_2-x_3+\frac{3}{4}}\nn
&&+4\frac{ x_1^2 \left(x_2^2+x_3^2\right)+x_2^2 x_3^2}{x_1-x_2+x_3+\frac{3}{4}}+4\frac{ x_1^2 \left(x_2^2+x_3^2\right)+x_2^2 x_3^2}{-x_1+x_2+x_3+\frac{3}{4}}\nn
&&+\frac{10252 \left(x_3^2 \left(x_1^3+x_2^3\right)+x_3^3 \left(-\left(x_1^2+x_2^2\right)\right)+x_1^2 x_2^2 (x_1+x_2)\right)}{9 (4 x_1+4x_2-4 x_3+3)^2}\nn
&&+\frac{2563 \left(x_1^3 \left(x_2^2+x_3^2\right)+x_1^2 \left(x_3^3-x_2^3\right)+x_2^2 x_3^2 (x_3-x_2)\right)}{36 \left(x_1-x_2+x_3+\frac{3}{4}\right)^2}\nn
&&+\frac{2563 \left(x_1^3 \left(-\left(x_2^2+x_3^2\right)\right)+x_1^2 \left(x_2^3+x_3^3\right)+x_2^2 x_3^2 (x_2+x_3)\right)}{36 \left(-x_1+x_2+x_3+\frac{3}{4}\right)^2}\bigg].
\ea
 Instead of $F(1,1,1)$, in the previous expression we use the amplitude given by $$f_{\rm NL}(\gamma_u/H,\gamma/H)=\frac{5}{6} F(1,1,1)/3P_\zeta^2$$ whose dependence on $\gamma$, $\gamma_u$ is shown in figure \ref{fig:Fv1}(b) for $c_{\rm sr}=\sqrt{1/3}$. It is suppressed for strong dissipation, as opposed to the $n^\mu$ term which grows as $f_{\rm NL}\sim \gamma_n$. The dependence can be approximated by the function
\beq
f_{\rm NL}=\frac{\gamma_u}{H} \frac{0.24 (\gamma/H)^{-0.26}}{1+0.038 (\gamma/H)^{0.60}},
\eeq
and it scales as $f_{\rm NL}\sim c_{\rm sr}^{-2}$ for a generic speed of sound of the fluid. 

Therefore, several possibilities appear. On the one hand, one could have high $\gamma$ and $\gamma_n$ but low $\gamma_u$ and in this case the contribution from this term to the three-point function is negligible and the result from \cite{diseft} dominates. On the other hand, one could have high $\gamma$ high $\gamma_u$ and $\gamma_n\simeq0$ and in this regime the strength of non-Gaussianities will be $u^\mu$ dominated and negligible. The specific relation between $\gamma$, $\gamma_u$ and $\gamma_n$ is arbitrary and one cannot say a priori which will be the case. 

This suppression for strong dissipation can be understood qualitatively as follows, considering expression \eqref{eq:vtermng} for the equilateral configuration. Recalling the properties of the clock's Green function, the $\eta'$ integral just picks up the time at which the modes freezed out, given by $\lambda_{\rm phys}=c_s/\sqrt{\gamma H}\gg c_s/H $. Therefore the relevant interactions occurred at an early time when the modes were well inside the horizon; the stronger the dissipation the earlier the interaction and the smaller the physical wavelength. In the case of the $n^\mu$ term the source is given by $\partial^i\pi \partial_i\pi=\partial_i\pi \partial_i\pi/a^2$, for the $u^\mu$ term it is $\partial_i u_L \partial_i\pi/a^2$. 

First of all, since the integral of $G_\gamma(0,y)$ gives $1$ and the change of variable to conformal time brings a $(kc_s\eta')^{-2}=(\omega/H)^{-2}$ factor, we can evaluate the integrand at freeze-out and estimate 
\beq
\frac{1}{H^2}\int d(kc_s \eta') \frac{G_\gamma(0,kc_s \eta')}{(kc_s\eta')^2}\sim \frac{1}{\omega^2}=\frac{1}{c_s^2k^2_{\rm phys}}
\eeq

 We consider first the $n^\mu$ term, in which the $(kc_s\eta')^2$ factor in the clock response cancels the $a^{-2}$ from the source giving the correct estimate 
\bea
f_{\rm NL} &\sim& \frac{\gamma_n}{H} \int \frac{G_\gamma(0,kc_s \eta')}{(kc_s\eta')^2} \frac{\lb \partial_i \zeta \zeta_f \rb}{a\lb \zeta_f^2\rb} \frac{ \lb \partial_i \zeta \zeta_f \rb }{a\lb \zeta_f^2\rb }\nn
& \sim& \frac{\gamma_n}{H}\frac{1}{c_s^2k^2_{\rm phys}} k^2_{\rm phys}\frac{\lb  \zeta \zeta_f \rb}{\lb \zeta^2\rb} \frac{ \lb  \zeta \zeta_f \rb }{\lb \zeta_f^2\rb }\sim \frac{\gamma_n}{Hc_s^2}
\ea

On the other hand, for the non-Gaussianity generated by $u^\mu$, using the fluid green function properties, the three-point function for the equilateral triangle is schematically 
\bea
f_{\rm NL} &\sim& \frac{\gamma_u}{H} \int \frac{G_\gamma(0,kc_s \eta')}{(kc_s\eta')^2} \frac{\lb \partial_i u_L \zeta_f \rb}{a \lb \zeta_f^2\rb} \frac{ \lb \partial_i \zeta \zeta_f \rb }{a \lb \zeta_f^2\rb } \nn
&\sim& \frac{\gamma_u}{H} \frac{1}{c_s^2k^2_{\rm phys}}\frac{\lb \partial_i u_L \zeta_f \rb}{a \lb \zeta_f^2\rb} \frac{ \lb \partial_i \zeta \zeta_f \rb }{a \lb \zeta_f^2\rb }\nn
&\sim&\frac{\gamma_u}{Hc_s^2} \frac{\lb u_L \zeta_f \rb}{ \lb \zeta_f^2\rb}\bigg|_{\text{at freeze-out}},
\ea 
where we used the fact that since the source of fluid perturbation dominates at freeze-out, the fluid response is nearly local. At this point if the dissipation is strong enough such that the time at which the interaction takes place the fluid perturbation is well inside its sound horizon then using the properties of $G_v$ gives the estimate 
\beq
f_{\rm NL}\sim\frac{\gamma_u}{Hc_{sr}^2} \left(\frac{\gamma}{H}\right)^{-1},
\eeq
which coincides with the estimate made in flat space for $\gamma\rightarrow\infty$. This was confirmed computing numerically the correlation $\lb u_L \zeta_f \rb$ at freeze-out as a function of $\gamma$ as shown in figure \ref{fig:upi}(b). The non-local behavior of the dissipation has the time-scale of the fluid's red-shift which is Hubble, and therefore there are corrections coming from the curvature of de Sitter. As explained in section \ref{sec:flat}, this correction is not negligible and tends to slightly increase the $f_{\rm NL}$ for increasing $\gamma/H$. 

This can be confirmed seeing how the strength of this interaction to the non-Gaussianities depends on the dissipation coefficient $\gamma$, fixed by the two-point function and the term $\gamma_u$. The dependence is shown in figure \ref{fig:Fv1}(c) where we plot the parameter $f^{\rm eq}_{\rm NL}$ from $\gamma_u u^\mu\partial_\mu\pi$ for radiation $c_{sr}^2=1/3$ as a function of $\gamma/H$. 
It can be seen that the strength is proportional to $\gamma_u$ but highly suppressed for high $\gamma$. 
\begin{figure}[t!]
\begin{center}
\subfigure[]{
\includegraphics[scale=0.4]{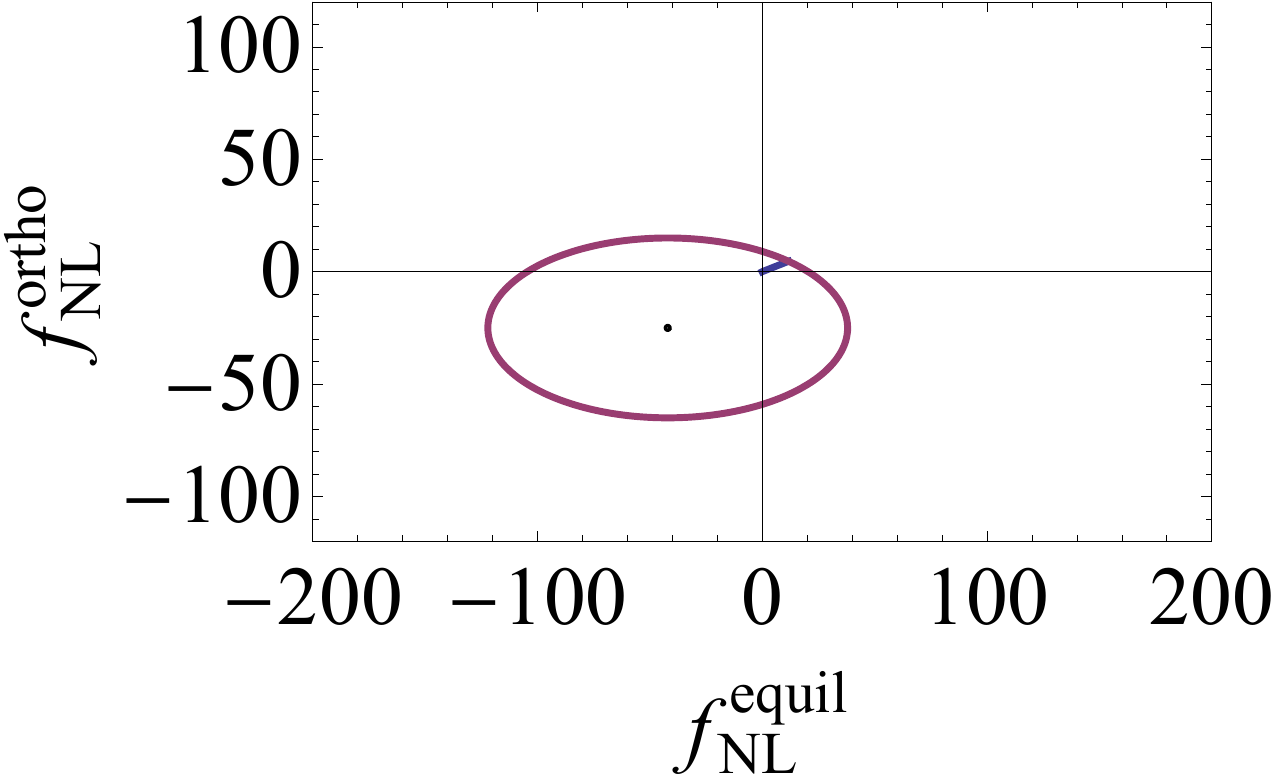}}
\subfigure[]{
\includegraphics[scale=0.42]{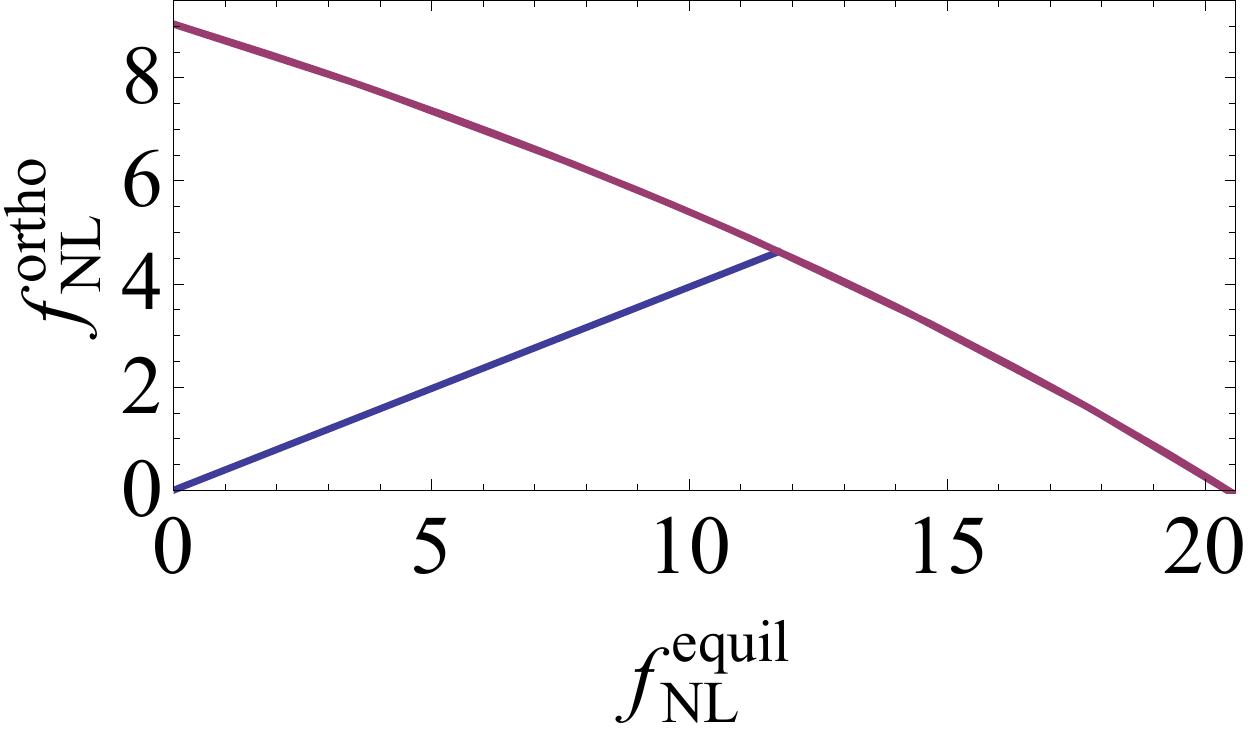}}
\caption{\label{fig:planck} \small\it (a) Plot of the pairs $(f_{\rm NL}^{\rm equil}, f_{\rm NL}^{\rm ortho})$ for $0<\gamma/H<\gamma_{\rm max}$ (blue line at the center). In red we show the boundary of the $99.7\%$ confidence region of the parameters. From the numerical calculation of $F(1,1,1)$ one concludes that $\gamma_{\rm max}\sim 10^5$ corresponds to the intersection between the blue and the red curve on the right panel. (b) Plot showing the curve corresponding to warm inflation in detail. }
\end{center}
\end{figure}
We will conclude in the next section that this term is the leading contribution to the three-point function coming from the interaction with the fluid. Then using the results of this section we can compute the parameters $f_{\rm NL}^{\rm equil}$, $f_{\rm NL}^{\rm ortho}$ and $f_{\rm NL}^{\rm loc}$ using the fudge factors or cosines. 

Recently thanks to the release of Planck results we know the $99. 7\%$ confidence region in ($f_{\rm NL}^{\rm equil}$,$f_{\rm NL}^{\rm ortho}$)-space \cite{planckng}. For the general case of the effective field theory of inflation coupled to ADOF with a preferred frame we can distinguish two important cases. If $\gamma\gg H$ and $\gamma_n \lesssim \gamma_u$ or $\gamma_n > \gamma_u$ the dominant contribution will be given by the local dissipation and the result does not change from the one explained on reference \cite{diseft}. 

On the other hand if $\gamma_n \simeq 0$ then the dominant contribution to the three-point function comes from the term studied in this section. Since the shape does not depend on $\gamma$ the points corresponding to this model in ($f_{\rm NL}^{\rm equil}$,$f_{\rm NL}^{\rm ortho}$)-space define a straight line starting at the origin (for $\gamma=0$) and defining a maximum $\gamma_{\rm max}$ such that $f_{\rm NL}( \gamma_{\rm max})$ is located in the boundary of the $99.7\%$ confidence region. 

The three-point function scales approximately as $f_{\rm NL} \sim 1/c_{sr}^2$ so for definiteness we assume we have radiation as in the case of warm inflation with $c_{sr}=\sqrt{1/3}$. Using the numerical results of this section we can compute the upper bound of the dissipation coefficient. As shown in figure \ref{fig:planck} it is given by the value of $\gamma$ corresponding to $f_{\rm NL}=28.5$ (obtained by dividing the value shown in the figure by the associated fudge factor). This is equivalent to the bound 
\beq
\gamma \leq 9.7\cdot 10^4 H.
\eeq
This bound is bigger than previous ones \cite{moss,planckng} because as we saw the three-point function increases much slower with increasing dissipation. For decreasing speed of sound of the fluid perturbation (even if the background equation of state does not change) the three-point function increases and therefore this upper bound of $\gamma/H$ gets reduced. 

Moreover, in \cite{moss,moss2} it was shown that there is a non-trivial three-point function in the squeezed triangle limit. In the full calculation done in this paper we see that this is not the case. In turn, the projection to the local shape we found $f_{\rm NL}^{\rm loc}$ is rather small compared to previous results. For radiation with $0<\gamma/H\lesssim 10^5$ it is given by 
\beq
0<f_{\rm NL}^{\rm loc}<1.3,
\eeq
which is clearly compatible with the Planck result $f_{\rm NL}^{\rm local}=2.7\pm5.8$. This combination of smaller local three-point function for a given equilateral configuration is more compatible with Planck results than previous approximate calculations.
\subsection{Squeezed Limit}
~~~~~Clearly from the shape shown above the squeezed limit of the three-point function vanishes, as opposed to what was found in \cite{moss}. This can be understood from the derivation of the consistency condition which is based on the ability of removing long wavelength modes by a re-scaling, making their effect disappear and giving a vanishing correlation with short modes. Of course this is only so in the limit of de Sitter and scale invariance and corrections to it will give suppressed contributions proportional to the slow-roll parameters, as explained in the Introduction. 

We think this is correct and indeed warm inflation-like models should give a trivial squeezed limit for the following reasons. First of all, if one goes to the unitary gauge the ADM parameters are given schematically by 
\bea
N&=&1+\frac{\dot{\zeta}}{H}+\frac{1}{2M_P^2}(\overline{\rho}_\calo+\overline{p}_\calo) \partial^{-2}\partial_i u^i+\ldots\rightarrow 1\\
N^i&=& -\frac{1}{a^2H}\partial_i\zeta+\partial_i \chi+\ldots \rightarrow 0
\ea
where $\partial^2\chi=\epsilon \dot{\zeta}+(1-c_{sr}^2)(2M_P^2H)^{-1}\delta\rho$, the dots represent terms of higher order in perturbations and the limit is taken for modes outside the horizon at late times. The contribution from the curvature perturbation goes to zero for the same reason that in single field inflation. The contribution from density and velocity fluctuation in the fluid goes to zero because they are sourced by inhomogeneities in the clock, either time or spatial gradients, which both go to zero after horizon exit. Therefore the only effect of a long wavelength mode of $\zeta$ is a rescaling of the spatial coordinates and can be removed by $x\rightarrow e^{\zeta_L}x$. The argument is the same as in \cite{consist1, consist2,diseftcon}.
\begin{figure}[t!]
\begin{center}
\includegraphics[scale=0.65]{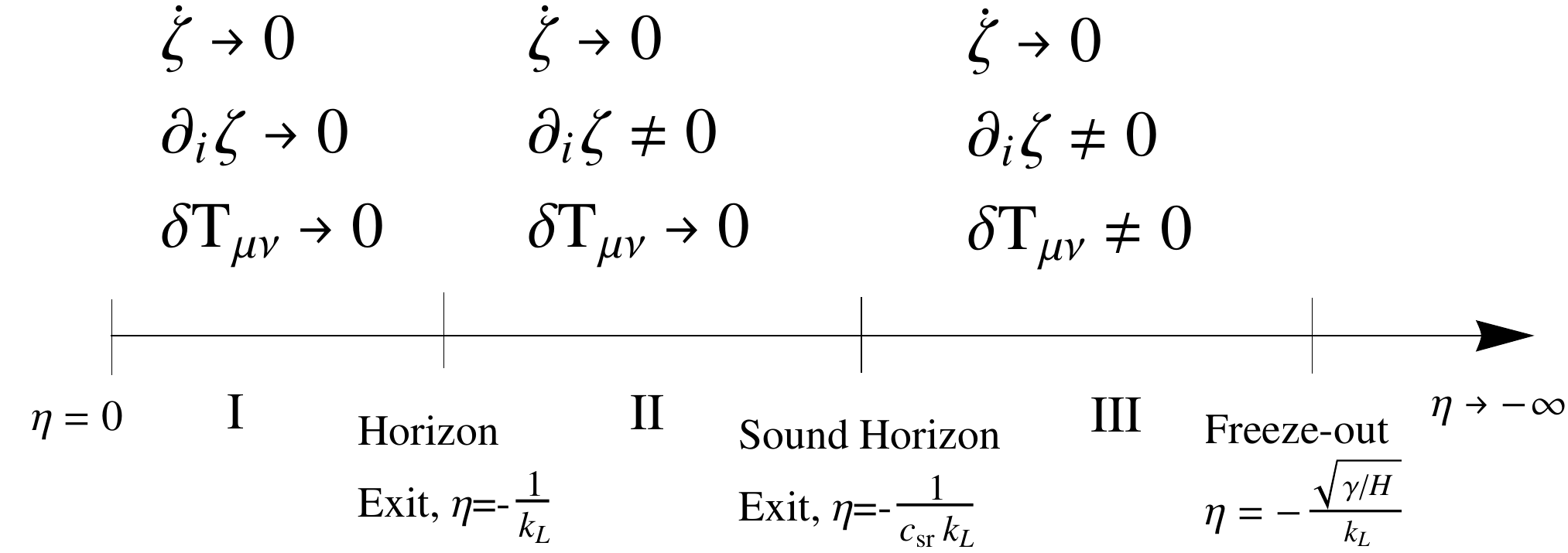}
\caption{\label{fig:timeline} \small\it The line indicates times ranging from the asymptotic past ($\eta\rightarrow-\infty$) to the asymptotic future ($\eta=0$). Three regions relevant to the effect of the long mode are shown. $\zeta$ is the curvature perturbation and $\delta T$ the perturbation of the fluid stress-energy tensor. (I) Far future, the mode already freezed-out, $\dot{\zeta}=0$, exited the sound horizon, $\delta T_{\mu\nu}=0$, and exited the Hubble horizon, $\partial_i \zeta=0$. (II) The mode already freezed-out, $\dot{\zeta}=0$, exited the sound horizon, $\delta T_{\mu\nu}=0$, but did not exit the Hubble horizon, $\partial_i \zeta\neq0$. (III) The mode already freezed-out, $\dot{\zeta}=0$ but did not exit neither the sound horizon, $\delta T_{\mu\nu}\neq0$ or the Hubble horizon, $\partial_i \zeta\neq0$.}
\end{center}
\end{figure}
Therefore since the non-Gaussianity studied above is a zeroth order effect in slow-roll there should not be any contribution to the three-point function in the squeezed limit and indeed is what we found. Terms contributing to the squeezed limit are the same as those studied in \cite{diseftcon} and thus the consistency condition is still valid in these systems. 

Before moving on, the models studied in this paper have distinct regimes according to the specific value of the ratio $k_L/k_S$ which are worth specifying. In figure \ref{fig:timeline} we show the history of the long mode. If the short mode freezes in region I ($k_S>\sqrt{\gamma/H}k_L$), which is the true squeezed limit, the argument above remains unchanged. If the short mode freezes in region II ($k_S>c_{sr}\sqrt{\gamma/H}k_L$) corrections appear due to the inhomogeneities in $\zeta$ but the fluid perturbation already decayed. Then the corrections are still mainly due to spatial curvature and as explained in the next paragraph $f_{\rm NL}\sim k_L^2$ so the transition from case I to II should be smooth. If the short mode freezes in region III ($k_S>k_L$) there are fluid perturbations and inhomogeneities in $\zeta$ so the arguments are no longer valid. 

In \cite{cremi,consist3} it was proven that in single-field models of inflation corrections to the squeezed limit scale as $(k_L/k_S)^2$. This is due to the fact that a long mode, to first non-trivial order, contribute as a source of curvature of order $k_L^2$ and therefore the squeezed limit is by the dependence of the power spectrum on the presence of spatial curvature. In the case considered in this paper of a $u^\mu\partial_\mu \pi$ term, this argument is still valid and $f_{\rm NL} \sim \calo(k_L^2)$. Near the squeezed limit, one can make a change of coordinate and go to the frame comoving with the fluid. This is possible because the difference between going to a frame with $N^i=0$ and $N=1$ and the fluid rest frame is sub-leading in slow roll. In that frame the effect of a long mode is spatial curvature and therefore the argument in \cite{consist3} is the same.

The behavior in quasi-single field inflation is different and the squeezed limit depends on the mass of the extra fields \cite{QSF}. In \cite{cremi} it is explained that this is due to the lack of scale invariance of the extra degrees of freedom. In this case for noise induced correlations the two point function of the fluid is indeed scale invariant and therefore the argument presented above is valid.  

In \cite{consist3} they argue that by continuity the behavior $\calo(k^2)$ should give the right parametric dependence of the three-point function strength. For example, for local dissipation, like reference \cite{diseft}, the wavelength at freeze-out scales as $k^2\sim \gamma$ and therefore $f_{\rm NL}\sim \gamma$, implying that $f^{n\partial \pi}_{\rm NL} \sim k^2 \calo(1)$. For our case, given the same assumptions, since $f^{\rm equi}_{\rm NL}$ is suppressed with respect to $\gamma$ this implies that $f^{u\partial \pi}_{\rm NL}\sim k^2 \calo( \gamma^{-1})$. Therefore near the squeezed limit the dominant contribution to order $k_L^2$ will come solely from the terms studied in \cite{diseft,diseftcon}. 

\subsection{Non-Gaussianities from the noise}
~~~~~Here we will describe the contribution from non-linearities in the noise kernel of the stochastic component of the ADOF. As explained above these corrections are not constraint by any symmetry or property or parameter of the background and may be there or not. We will study it anyways because of the possible presence of FDT and depending on the microscopic theory the may not be arbitrary. We define:
$$\beta_\nu \equiv \frac{\partial\log\nu_\calo}{\partial\rho/3(\rho+p)},$$
so that the two-point function for the stochastic source is now
\beq
\lb \delta\calo_S(t,\mathbf{k})\delta\calo_S(t',\mathbf{q})\rb=\frac{\nu_\calo}{\sqrt{-g}}\left(1+\beta_\nu \delta_\mathbf{k}(t)\right)\delta(t-t')\delta(\mathbf{k}+\mathbf{q}).
\eeq
Since $\delta_\mathbf{k} \sim \int G_R \partial^2\pi\sim \int G_R\int G_\gamma \delta\calo_S$ this will give a non-vanishing three-point function for $\pi$. As shown in \cite{diseftcon}, the effect of changing the kernel can be imitated by adding a new term in the equation of motion of the form 
\beq
-\beta_\nu \frac{1}{2N_c} \delta({\bf x},\eta) \delta\calo_{\mathcal{S}}({\bf x},\eta).
\eeq
For example, in the case of a canonical scalar field coupled to radiation, from the fluctuation-dissipation theorem one concludes that $\beta_\nu=1$.

\begin{figure}[t!]
\begin{center}
\subfigure[]{
\includegraphics[scale=0.44]{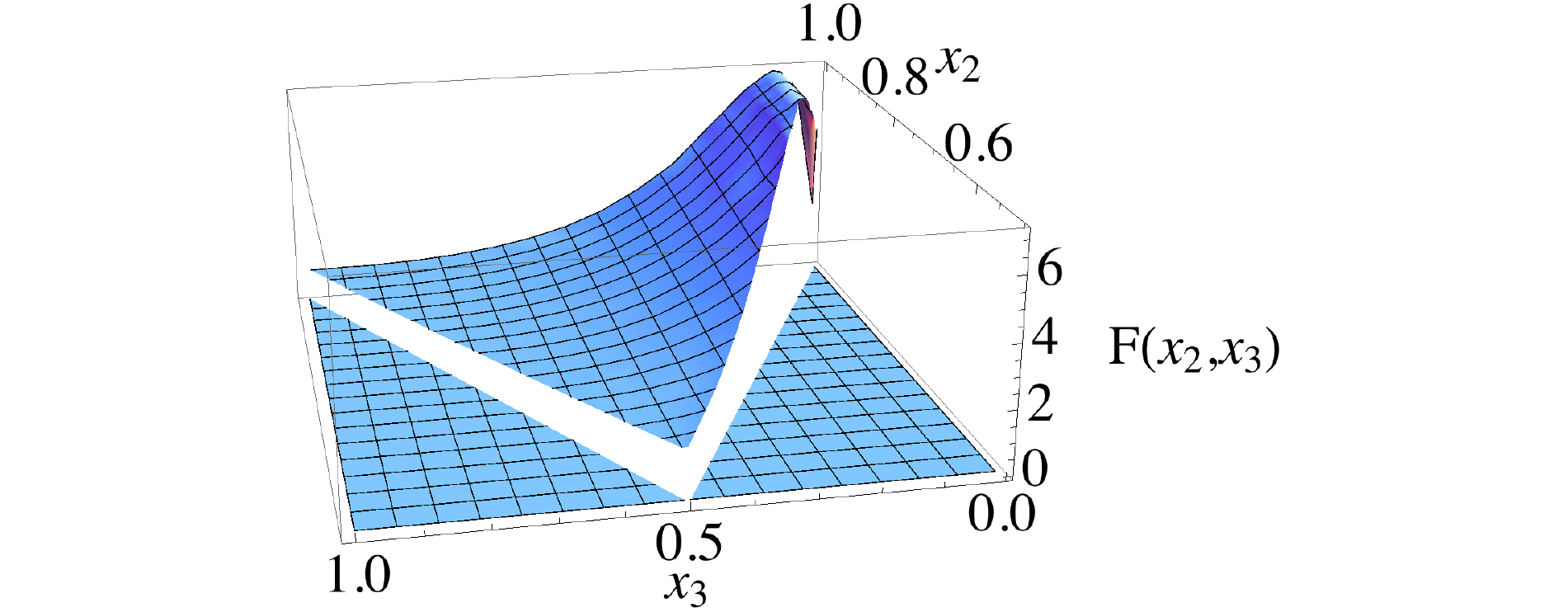}}
\subfigure[]{
\includegraphics[scale=0.4]{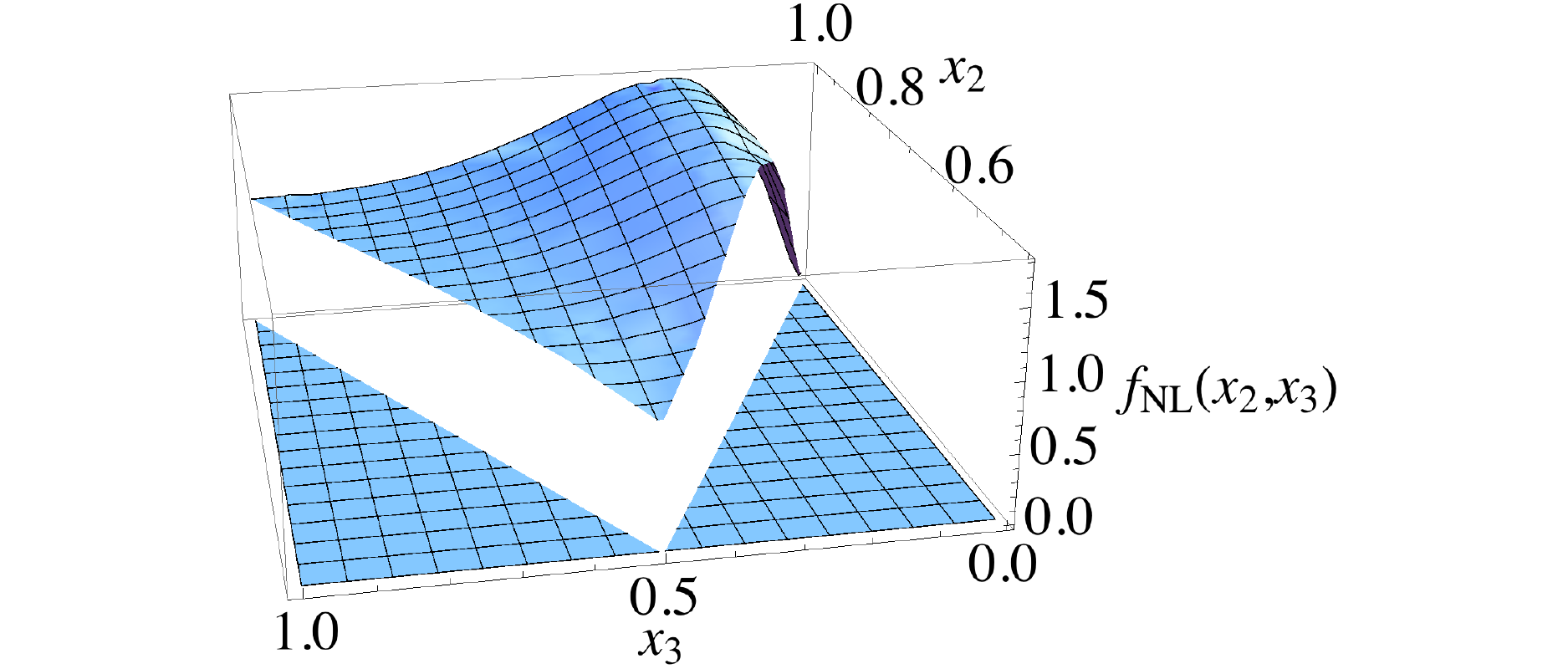}}
\caption{\label{fig:noise1} \small\it (a) Plot of the three point function $F(x_2,x_3)$ for a dissipation of $\gamma/H=40$ and a fluid with sound speed $c_{sr}=\sqrt{1/3}$ (b) Plot of the dependence of the $f_{\rm NL}$ parameter with respect to the triangle shape. One can see that near the squeeze limit $f_{\rm NL}\rightarrow 0$.}
\end{center}
\end{figure}

Following the same steps as in the last section, one can write the contribution to the three-point function as  
\beq\label{eq:noisetermng}
F(k_1,k_2,k_3)=\beta_\nu \sum_{\rm perm.}\frac{1}{2N_c}  \int d(k_1c_s\eta') \frac{G_\gamma(0,k_1 c_s |\eta'|)}{(k_1c_s\eta')^2}\lb \delta_{\mathbf{k}_2}(\eta') \zeta_{\mathbf{k}_2f}\rb \lb\delta\calo_{\mathcal{S}}(k_3,\eta')\zeta_{\mathbf{k}_3f}\rb.
\eeq
The explicit expression in terms of the Green functions of the clock and fluid is given in the appendix. We again discard the terms sourced by time derivatives of the clock since they are subleading. The shape of the non-Gaussianity generated by this effect was computed numerically and the result is shown in figure \ref{fig:noise1}. Once again, a quantitative way of describing this shape is through the cosines and fudge factors with a stablished shape, which we take to be again the equilateral, orthogonal and local. 
For this contribution to the three-point function we obtain 
\bea
&&\cos\theta_{\rm equil}\simeq0.58,\\
&&\cos\theta_{\rm ortho}\simeq-0.57, \\
&&\cos\theta_{\rm local}\simeq0.96.
\ea
These cosines were computed for $\gamma=40H$ and $c_{sr}=\sqrt{1/3}$. As a function of $\gamma$ the cosines don't change an appreciable amount but the peak seen in figure \ref{fig:noise1}(b) gets displaced to lowers values of $x_3$. This is because for $k_3/k_2\lesssim\sqrt{H/\gamma}$ the long mode is well outside the horizon when the interaction is taking place (freeze-out of the short modes). Then, the source of density perturbations in the fluid decays and therefore the $f_{\rm NL}$ drops to $0$ in the squeezed limit.

As we did before, instead of the cosines one could give the fudge factors writing the shape in the following way
\beq
\frac{F(x_1,x_2,x_3)}{F(1,1,1)}\simeq1.23F_{\rm equil}-0.91F_{\rm ortho}+0.56F_{\rm local}+\ldots,
\eeq
where the dots represent other possible shapes not considered here. 
\begin{figure}[t!]
\begin{center}
\includegraphics[scale=0.6]{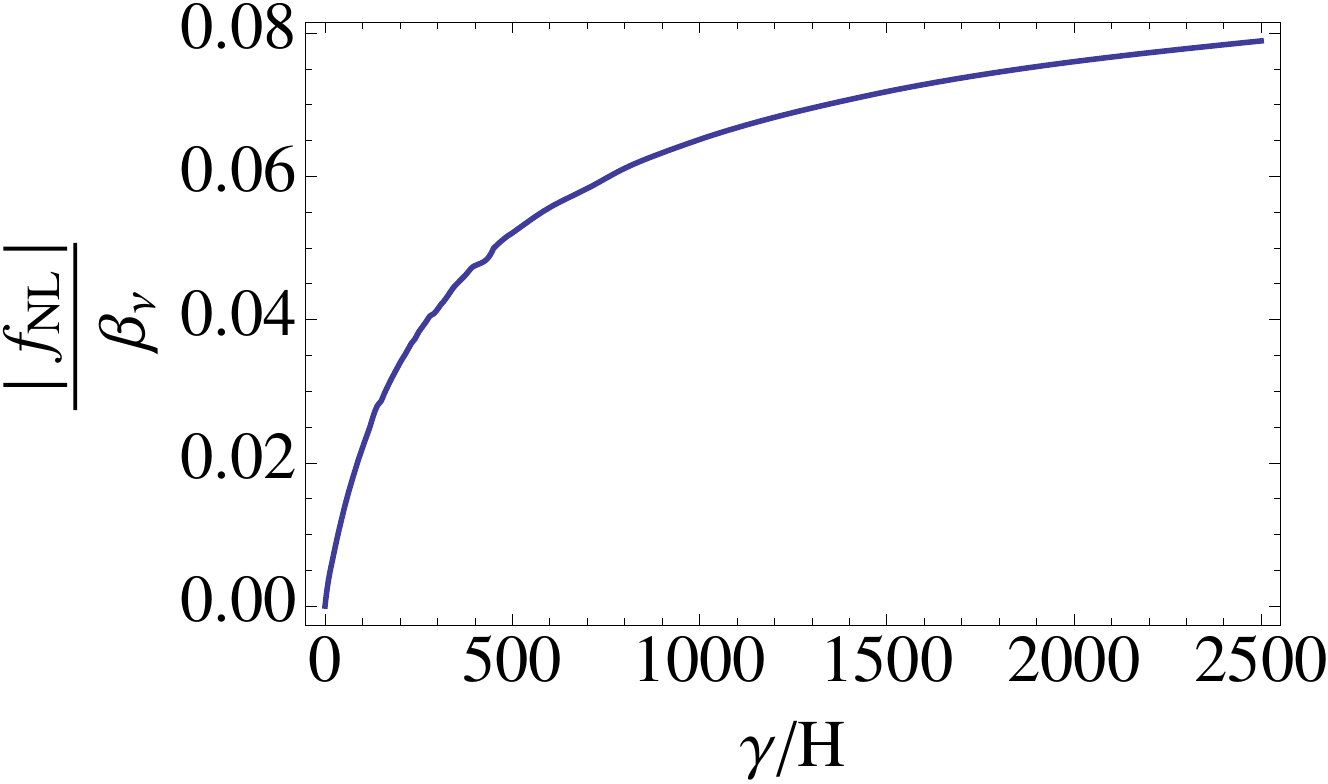}
\caption{\label{fig:noise_amp} \small\it Plot of the parameter $|f_{\rm NL}|=-f_{\rm NL}$ for a fluid with sound speed $c_{sr}=\sqrt{1/3}$ as a function of $\gamma/H$.}
\end{center}
\end{figure}
Regarding the strength of this contribution, we measure it through the parameter $f_{\rm NL}=\frac{5}{6} F(1,1,1)/P_\zeta^2$ like in the previous cases. This quantity is shown in figure \ref{fig:noise_amp}. To understand the result one can follow the same reasoning as in the previous section 
\bea
f_{\rm NL} &\sim&\beta_\nu \int \frac{G_\gamma(0,kc_s \eta')}{(kc_s\eta')^2} \frac{\lb \delta(\eta') \zeta_f \rb}{ \lb \zeta_f^2\rb} \frac{ \lb N_c^{-1}\delta\calo_{\mathcal{S}}(\eta') \zeta_f \rb }{ \lb \zeta_f^2\rb } \nn
&\sim&\beta_\nu \frac{1}{c_s^2k^2_{\rm phys}} \frac{\lb \delta(\eta') \zeta_f \rb}{ \lb \zeta_f^2\rb} \frac{\nu_\calo}{N_c^{2}c_s^3} \sqrt{\frac{H}{\gamma}} \frac{\gamma}{H} \frac{ 1}{ \lb \zeta_f^2\rb }\nn
&\sim&\beta_\nu \frac{\lb \delta(\eta') \zeta_f \rb}{ \lb \zeta_f^2\rb}\bigg|_{\text{at freeze-out}},
\ea 
where we used the fact that for strong dissipation $$\lb \zeta_f^2\rb \sim \frac{\nu_\calo H^2}{c_s(c_sN_c)^2} \sqrt{\frac{H}{\gamma}}.$$ From the response of density fluctuations shown in figure \ref{fig:gra} one can see that the amplitude of fluctuations for sources produced in the far past is nearly independent of the time the are produced. Therefore, since increasing $\gamma$ is the same as considering fluctuations being produced earlier in time, the strength of this contribution to the $f_{\rm NL}$ reaches a constant value for strong dissipation. 

Regarding the shape of this contribution. Starting from the equilateral triangle configuration, the three-point function starts increasing as one approaches the squeezed triangle. This is because the effect of the clock perturbation on the fluid starts acting at freeze-out (local in the equilateral since every mode freezes simultaneously) and then the source accumulates starting from freeze-out of the long mode until the time when the density perturbation interacts, at freeze-out of the short mode. Eventually it drops because the density perturbation starts acting after the mode exited the Hubble horizon $k_L/k_S<\sqrt{\gamma/H}$. Then the three-point function starts decaying because neither the source of $\delta\rho$ or $\partial^2\zeta$ is efficient nor sound waves are present because if $c_{sr}<1$ the mode has already exited the sound horizon. Finally, it vanishes in the squeezed limit.  

One can see that this effect is subleading with respect to the contribution to the three-point function coming from $u^\mu\partial_\mu\pi$ and in the strong dissipation regime it will be negligible\footnote{Having a $\beta_\nu\gg1$ enhances the signal but most models rely on a fluctuation-dissipation theorem that constraints it to be $\beta_\nu\simeq1$}. This same conclusion is reached in the dissipative EFT of inflation without sound waves: the leading term is always the one coming from the non-linear realization of coordinate invariance in the dissipative term. Therefore the inclusion of this term is irrelevant for the comparison of warm inflation with Planck data.

\section{Conclusions}\label{sec:conclusions}
~~~~~We have studied how a fluid behaves in an inflationary background in the context of dissipative EFT of inflation with a non-vanishing expectation value of a time-like four vector defining two preferred frame: the one where the clock is homogeneous and the one comoving with the fluid. We used a perfect fluid with a speed of sound not necessarily equal to the background value and studied its evolution, although generalizing the calculation is straightforward and no qualitative changes to our conclusions are expected. 

We observed that fluctuations are driven by inhomogeneities of the clock and that therefore they decay at late times when the mode exits the horizon, making both preferred clocks coincide at late time. This is critical in understanding why the shape of the three-point function resembles the one with a single preferred clock and does not develop a non-trivial squeezed limit as opposed to what was found in \cite{moss,moss2}.

 This must be the case since the fluid is sensitive to gradients and time derivatives and so a constant, homogeneous $\pi$ perturbation can be gauged away. Moreover, because the fluid is sourced by the clock itself a non-trivial squeezed limit would require giving the $\pi$ field a mass and therefore giving $\zeta$ a time evolution outside the horizon. Even if that is not the case, it could mean having a second preferred clock different from the one driving inflation. Since this does not happen this models of two clocks that coincide at late times resemble more the dissipative case than the multi-field case and this is reflected on the three-point function. 
 
 We also computed the three-point function numerically checking that this interpretation is correct and generalizing it to a generic clock and fluid but with warm inflation in the back of our minds. Another characteristic of the non-local interaction mediated by sound waves of this second clock is the fact that $|f_{\rm NL}|\sim 1/c_{sr}^{2}$ (with $c_{sr}$ the speed of sound in the fluid) in the strong dissipation regime; while for a single clock is $|f_{\rm NL}|\sim \gamma/Hc_{s}^{2}$. Therefore if the latter component is present it will always dominate over the former in the strong dissipation regime. If the former is the only contribution then the three-point function will not be enhanced unless the sound speed of the fluid is small. 
 
 In \cite{planckng} it was noticed that the correlation between the measured non-Gaussian signal with the quasi-local three-point function found in \cite{moss} is rather small since the data favors shapes closer the equilateral or orthogonal. Here we computed the $f^{\rm equi}_{\rm NL}$ and $f^{\rm ortho}_{\rm NL}$ parameters for warm inflation and checked that they can be well inside the confidence region. This shows that using our results the warm inflationary scenario is still at least phenomenologically a viable option as a means of generating the non-Gaussianities measured (or bounded) by the satellite.

\subsection*{Acknowledgments}

We thank Diana Lopez Nacir for useful discussion and for sharing a template to fit the non-gaussian shape. M. Z. is supported in part by the National Science Foundation grants PHY-0855425, AST-0907969, PHY-1213563 and by the David \& Lucile Packard Foundation.

\begin{appendix}
\section*{Appendix}
\section{Calculation of non-Gaussianities\label{app:ng}}
In this appendices we will go through the details of the computation of the three-point function for the different sources of interactions. In previous sections we wrote it in terms of correlations between the clock and the fluid to make the physics more transparent; here we will give the precise expressions that we used to do the numerical integrations. The procedure follows the presentation of \cite{diseft,diseftcon}.

\subsection{Three-Point function from the $u^\mu\partial_\mu\pi$ term}
In this section we will compute the three-point function coming from the $u^\mu\partial_\mu\pi$ term in the equation of motion of the clock. To do this calculation one has to expand the clock as $\pi=\pi_1+\pi_2+\ldots$, where $\pi_1$ represents the solution to the equation of motion retaining only linear terms, equation \eqref{pi1}, and $\pi_2$ is the leading order correction coming from the particular interaction considered in this section mediated by the fluid. Using linear response theory this last contribution is given by 
\bea
\pi_2&=&\frac{kc_s}{H^2}\int d\eta' \frac{G_\gamma(kc_s|\eta|,kc_s|\eta'|)}{(kc_s|\eta'|)^2}\gamma_u[u_1^i\ast\partial_i\pi_1]_\mathbf{k}\nn
&=&\frac{kc_s}{H^2}\int d\eta' \frac{G_\gamma(kc_s|\eta|,kc_s|\eta'|)}{(kc_s|\eta'|)^2} \nn
&&\times\frac{\gamma_u H^2}{c_{sr}}\int \frac{d^3q}{(2\pi)^3} \int d\eta'' \hat{{\bf q}}\cdot({\bf k-q}) \tilde{G}_V(c_{sr}q\eta',c_{sr}q\eta'') \pi_1({\bf q},\eta'')  \pi_1({\bf k-q},\eta'),
\ea
where $\tilde{G}_V(z,y)=3zG_V(z,y)$. Now we can compute the three-point function for curvature perturbations $\zeta\simeq -H\pi$ in the limit of late times $\eta\rightarrow0$ as the cyclic sum of correlations between two first order fluctuations and a second order one:
\bea\label{eq:3point}
\lb \zeta_{\mathbf{k}_1}\zeta_{\mathbf{k}_2}\zeta_{\mathbf{k}_3}\rb &=&-H^3\sum_{\rm cyclic} \lb {\pi_1}(\mathbf{k}_1,0){\pi_1}(\mathbf{k}_2,0){\pi_2}(\mathbf{k}_3,0)\rb\nn
&\equiv&(2\pi)^3\delta(\sum_ik_i)F(k_1,k_2,k_3).
\ea
 Using the explicit expressions for the clock perturbation we arrive at the expression
\bea
F(k_1,k_2,k_3)&=&\frac{\gamma_u}{H}  \hat{{\bf k}}_1\cdot\hat{{\bf k}}_2\frac{k_3 k_2c_sH^4}{c_{sr}}\int_{\eta_0}^\eta d\eta' g_\gamma(0,k_3c_s\eta')\int^{\eta'}_{\eta_0}d\eta''   \tilde{G}_V(c_{sr}k_1\eta',c_{sr}k_1\eta'')\nn
&&~~~\times \lb \pi({\bf k}_1)\pi(-{\bf k}_1,\eta'')\rb \lb \pi({\bf k}_2) \pi(-{\bf k}_2,\eta')\rb+\text{perm.}.
\ea
From this expression is clear why the terms with $\dot{\pi}$ are negligible since one has to replace the first correlation in the equation above with $\lb \pi_f \dot{\pi}\rb$ which is suppressed for strong dissipation according to the behavior seen in figure \ref{fig:pi1} and already noticed in \cite{diseft}. Moreover, those terms do not contribute to the squeezed limit neither so we can neglect them and take the expression above as an accurate approximation to the full three-point function coming from $u^\mu\partial_\mu\pi$.

Using momentum conservation the dot product is $2{\bf k}_1\cdot{\bf k}_2=(k_3^2-k_1^2-k_2^2),$ and the shape can be analyzed as in the usual case in terms of $k_1,k_2,k_3$. The correlation $\lb \pi_f \pi(\eta)\rb$ can be computed exactly, although the exact expression is not very illuminating. To extract the power spectrum at late times, it can be written as $H^2k^3\lb \pi({\bf k})\pi(-{\bf k},\eta)\rb = \Delta_\zeta \mathcal{F}_\gamma(c_sk\eta)$ where $\mathcal{F}(c_sk\eta)$ is a smooth function that satisfying $\mathcal{F}(c_sk\eta\rightarrow0)=1$ and $\mathcal{F}(c_sk\eta\gtrsim z_\star)=0$ and it is explicitly given by 
\beq
 \mathcal{F}_\gamma(z)=\frac{\pi  \Gamma(4 \nu-2)}{2^{8 \nu-9} (\nu -1) \Gamma(\nu-1)^2 \Gamma(\nu)^2} \int_{z}^{\infty} dy~ G_\gamma(0,y)G_\gamma(z,y),
\eeq
which is equal to the right hand side of equation \eqref{pipif}.

Following the notation introduced in \cite{shape} we will rewrite the momentum dependence in terms of the variables $x_i=k_i/k$, where in principle $k$ is some arbitrary scale that we will chose to be $k_1$, and renaming $F(x_1,x_2,x_3)=k^6 F(k_1,k_2,k_3)$ the three-point function can be written as 
\bea
F(x_1,x_2,x_3)&=&\frac{\gamma_u}{H}\Delta_\zeta^2  \frac{(x_3^2-x_1^2-x_2^2)x_3}{2x_1^4x_2^3}\frac{c_s}{c_{sr}}\int_0^\infty dy~g_\gamma(0,x_3c_sy)\nn
&&\times\int^\infty_y dz~\tilde{G}_V(c_{sr}x_1y,c_{sr}x_1z) \mathcal{F}_\gamma(c_sx_1z) \mathcal{F}_\gamma(c_sx_2y) +\text{ perm.}
\ea
Using the expression above we computed this quantity numerically and we present the result showing $F(x_2,x_3)=x_2^2x_3^2F(1,x_2,x_3)$ as done in \cite{shape}. The only relevant region displayed in $x_2,x_3$-space to avoid over-counting triangle configurations is $1-x_2\leq x_3 \leq x_2$. As opposed to the result found in \cite{moss} the shape has a trivial squeezed limit (at least to zeroth order in slow-roll) and therefore this term does not compromise the consistency condition derived for dissipative EFT of inflation with a preferred clock.

Using the expression for $F(x_1,x_2,x_3)$ we can compute the parameter $f_{\rm NL}$ as a function of the triangle as 
\bea
f_{\rm NL}(x_1,x_2,x_3)&=&\frac{5}{6}\frac{x_1^3x_2^3x_3^3}{x_1^3+x_2^3+x_3^3}\frac{\gamma_u}{H}  \frac{(x_3^2-x_1^2-x_2^2)x_3}{2x_1^4x_2^3}\frac{c_s}{c_{sr}}\int_0^\infty dy~g_\gamma(0,x_3c_sy)\nn
&&\times\int^\infty_y dz~\tilde{G}_V(c_{sr}x_1y,c_{sr}x_1z) \mathcal{F}_\gamma(c_sx_1z) \mathcal{F}_\gamma(c_sx_2y) +\text{ perm.}
\ea

For higher dissipation or different sound speeds the shape is very similar although the amplitude changes. The strength of this contribution is characterized by the bispectrum for the equilateral triangle $f_{\rm NL}^{\rm eq}$ and it is given by $f_{\rm NL}(1,1,1)$.

\subsection{Contribution from $\rho$ dependence of $\nu_\calo$}

The procedure now is equivalent to the one of the previous section. The only difference now is the non-linear term in the equation of motion sourcing $\pi_2$ which is  
\beq
-\beta_\nu \frac{1}{2N_c} \delta({\bf x},\eta) \delta\calo_{\mathcal{S}}({\bf x},\eta).
\eeq
The second order contribution to the clock perturbation is, using expression \eqref{deltaT} to write $\delta(k,\eta)$ in terms of $\pi_1$, given by
\bea
\pi_2(k,\eta)&=&\frac{kc_s}{H^2}\int d\eta' \frac{G_\gamma(kc_s|\eta|,kc_s|\eta'|)}{(kc_s|\eta'|)^2}\left(-\beta_\nu \frac{1}{2N_c} [\delta\ast \delta\calo_{\mathcal{S}}]_{\mathbf{k}}(\eta')\right)\nn
&=&-\frac{kc_s}{2N_cH c_{sr}}\int_{\eta_0}^{\eta} d\eta' \frac{G_\gamma(kc_s|\eta|,kc_s|\eta'|)}{(kc_s|\eta'|)^2}\int \frac{d^3\mathbf{q}}{(2\pi)^3}\nn
&&~~\times
  \int^{\eta'}_{\eta_0} d\eta'' q G_R(c_{sr}q|\eta'|,c_{sr}q|\eta''|)\pi_1(\mathbf{q},\eta'')
\delta\calo_S(\mathbf{k}-\mathbf{q},\eta').
\ea
Again, we neglect the contribution from terms sourced by time derivatives. Writing the three-point function of $\zeta\simeq-H\pi$ as a sum over cyclic permutations of two $\pi_1$ and one $\pi_2$ gives 
\bea
F(k_1,k_2,k_3)&=&\sum\beta_\nu \frac{k_3c_sH}{2N_c}\int d\eta' g_\gamma(0,k_3c_s|\eta'|)\lb \pi_{\mathbf{k}_1f}\delta\calo_S(\mathbf{k}_1,\eta')\rb \lb \delta_{\mathbf{k}_2}(\eta')\pi_{\mathbf{k}_2f}\rb\nn
&=&\sum\beta_\nu \frac{k_3c_sH^3}{2N_c}\int d\eta' g_\gamma(0,k_3c_s|\eta'|)\frac{\nu_\calo k_1c_s}{N_c}g_\gamma(0,k_1c_s|\eta'|)(\eta')^4\lb \delta_{\mathbf{k}_2}(\eta')\pi_{\mathbf{k}_2f}\rb\nonumber
\ea
\bea
&=&\sum\beta_\nu \frac{\nu_\calo k_3k_1c^2_sH^3}{2N^2_c}\int d\eta' g_\gamma(0,k_3c_s|\eta'|)g_\gamma(0,k_1c_s|\eta'|)(\eta')^4\nn
&&~~~~\times\frac{k_2H}{c_{sr}} \int d\eta'' G_R(k_2c_{sr}|\eta'|,k_2c_{sr}|\eta''|) \lb \pi_{\mathbf{k}_2}(\eta'')\pi_{\mathbf{k}_2f}\rb\nn
&=&\sum\beta_\nu\frac{1}{2}\frac{\nu_\calo^2H^4c_s^4}{N_c^4c_{sr}}k_3k_1k^3_2\int d\eta' g_\gamma(0,k_3c_s|\eta'|)g_\gamma(0,k_1c_s|\eta'|)(\eta')^4\nn
&&~~~~\times \int d\eta'' G_R(k_2c_{sr}|\eta'|,k_2c_{sr}|\eta''|)\nn
&&~~~~\times\int d\tilde{\eta} (\tilde{\eta})^4 g_\gamma(0,k_2c_s|\tilde{\eta}|)g_\gamma(k_2c_s |\eta''|,k_2c_s|\tilde{\eta}|).
\ea
The expression above can be simplified using the expression for $\Delta_\zeta$ and written in terms of $x_i$ giving the following result 
\bea
F(x_1,x_2,x_3)&=&\sum\beta_\nu\frac{\Delta_\zeta^2}{2}\frac{c_s^{10}}{c_{sr}}x_1x^3_2x_3\left(\frac{16^{\frac{\gamma}{H}}(\frac{\gamma}{H}+1)^3\Gamma(\frac{\gamma+H}{2H})^4}{\pi\Gamma(\frac{2\gamma}{H}+4)}\right)^{-2}\nn
&&~\times\int_{0}^{\infty} dzz^4g_{\gamma}(0,c_sx_3 z) g_{\gamma}(0,c_sx_1z) \int^{y}_{z} dw G_r(c_{sr}x_2z,c_{sr}x_2w)\nn
&&~\times\int_{0}^{\infty} dy y^4g_{\gamma}(0,c_sx_2y)g_{\gamma}(c_sx_2w,c_sx_2y).
\ea
Repeating the same steps the non-linear parameter is given by
\bea
f_{\rm NL}(x_1,x_2,x_3)&=&\sum\beta_\nu\frac{5}{12}\frac{x_1^3x_2^3x_3^3}{x_1^3+x_2^3+x_3^3}\frac{c_s^{10}}{c_{sr}}x_1x^3_2x_3\left(\frac{16^{\frac{\gamma}{H}}(\frac{\gamma}{H}+1)^3\Gamma(\frac{\gamma+H}{2H})^4}{\pi\Gamma(\frac{2\gamma}{H}+4)}\right)^{-2}\nn
&&\int_{0}^{\infty} dzz^4g_{\gamma}(0,c_sx_3 z)g_{\gamma}(0,c_sx_1z)  \int^{y}_{z} dw G_r(c_{sr}x_2z,c_{sr}x_2w)\nn
&&\int_{0}^{\infty} dy y^4g_{\gamma}(0,c_sx_2y)g_{\gamma}(c_sx_2w,c_sx_2y).
\ea
Using these expressions we computed numerically the three-point function. 

\end{appendix}

 \begingroup\raggedright\endgroup

\begin{thebibliography}{10}
 
 \bibitem{diseft} D.~Lopez Nacir, R.~A.~Porto, L.~Senatore and M.~Zaldarriaga,
  ``Dissipative effects in the Effective Field Theory of Inflation,''
  JHEP {\bf 1201}, 075 (2012)
  [arXiv:1109.4192 [hep-th]].
 
   \bibitem{multieft}  L.~Senatore and M.~Zaldarriaga,
  ``The Effective Field Theory of Multifield Inflation,''
  JHEP {\bf 1204}, 024 (2012)
  [arXiv:1009.2093 [hep-th]].
  
   \bibitem{warm} 
  A.~Berera, ``Warm inflation,''
  Phys.\ Rev.\ Lett.\  {\bf 75}, 3218-3221 (1995).
  [astro-ph/9509049]. ``Thermal properties of an inflationary universe,''
  Phys.\ Rev.\  {\bf D54}, 2519-2534 (1996)[hep-th/9601134]. ``The warm inflationary universe,''
  Contemp.\ Phys.\  {\bf 47}, 33 (2006)
  [arXiv:0809.4198 [hep-ph]].  A.~Berera and L.~Z.~Fang,
  ``Thermally induced density perturbations in the inflation era,''
  Phys.\ Rev.\ Lett.\  {\bf 74}, 1912 (1995)
  [arXiv:astro-ph/9501024].
   A.~Berera, I.~G.~Moss and R.~O.~Ramos,
  ``Warm Inflation and its Microphysical Basis,''
  Rept.\ Prog.\ Phys.\  {\bf 72}, 026901 (2009)
  [arXiv:0808.1855 [hep-ph]]. M.~Bastero-Gil, A.~Berera, ``Warm inflation model building,''
  Int.\ J.\ Mod.\ Phys.\  {\bf A24}, 2207-2240 (2009). [arXiv:0902.0521 [hep-ph]].
 
   \bibitem{moss} I.~G.~Moss and C.~Xiong,
  ``Non-Gaussianity in fluctuations from Warm Inflation,"
  JCAP {\bf 0704}, 007 (2007)
  [arXiv:0701302 [astro-ph]].

\bibitem{moss2} I.~G.~Moss and T.~Yeomans,
  ``Non-Gaussianity in the strong regime of  Warm Inflation,"
  [arXiv:1102.2833 [astro-ph]].
 
 
 \bibitem{maldacena}  J.~M.~Maldacena, ``Non-Gaussian features of primordial fluctuations in single field inflationary models,'' JHEP {\bf 0305}, 013 (2003)
  [arXiv:astro-ph/0210603].

\bibitem{consist1} P.~Creminelli and M.~Zaldarriaga,
  ``Single field consistency relation for the 3-point function,''
  JCAP {\bf 0410}, 006 (2004)
  [astro-ph/0407059].

\bibitem{consist2}
  C.~Cheung, A.~L.~Fitzpatrick, J.~Kaplan and L.~Senatore, 
  ``On the consistency relation of the 3-point function in single field inflation,''
  JCAP {\bf 0802}, 021 (2008)
  [arXiv:0709.0295 [hep-th]].
    
      \bibitem{diseftcon} D.~Lopez Nacir, R.~A.~Porto and M.~Zaldarriaga,
  ``The consistency condition for the three-point function in dissipative single-clock inflation,''
  JCAP {\bf 1209}, 004 (2012)
  [arXiv:1206.7083 [hep-th]].
  
    
       \bibitem{planckinf} P. Ade et al. (Planck Collaboration) (2013), 1303.5082.
   \bibitem{planckng} P. Ade et al. (Planck Collaboration) (2013), 1303.5084.
    \bibitem{eft1}   C.~Cheung, P.~Creminelli, A.~L.~Fitzpatrick, J.~Kaplan and L.~Senatore,
  ``The Effective Field Theory of Inflation,''
  JHEP {\bf 0803}, 014 (2008)
  [arXiv:0709.0293 [hep-th]].
  
  \bibitem{QSF} X.~ Chen and Y.~Wang,
  ``Quasi-single field inflation and non-Gaussianities,''
  JCAP {\bf 1009}, 027 (2010)
  [arXiv:0911.3380 [hep-th]].
  
  
    \bibitem{calzetta} E. Calzetta and B. L. Hu, ``Nonequilibrium Quantum Field Theory," Cambridge University Press (2008).
    
     \bibitem{trapped} D.~Green, B.~Horn, L.~Senatore and E.~Silverstein, ``Trapped Inflation," Phys.\ Rev.\  D {\bf 80}, 063533 (2009)
  [arXiv:0902.1006 [hep-th]].
   \bibitem{wistability}
   I.~G.~Moss and C.~Xiong,
  ``On the Consistency of Warm Inflation,"
  JCAP {\bf 0811}, 023 (2008)
  [arXiv:0808.0261 [astro-ph]].

   \bibitem{fluidos} S.~Endlich, A.~Nicolis, R.~Rattazzi and J.~Wang, 
``The quantum mechanics of perfect fluids," 
JHEP {\bf 1104}, 102 (2011)
  [arXiv:1011.6396 [hep-th]].
   
   \bibitem{weinberg} S.~Weinberg, 
``Cosmology", Oxford University Press, USA (2008).


     \bibitem{shape} D.~Babich, P.~Creminelli and M.~Zaldarriaga, 
   ``The shape of non-Gaussianities,"
   JCAP {\bf 0408}, 009 (2004)
   
  \bibitem{cremi} 
   P.~Creminelli, G.~D'Amico, M.~Musso and J.~Norena,
  ``The (not so) squeezed limit of the primordial 3-point function,''
  JCAP {\bf 1111}, 038 (2011)
  [arXiv:1106.1462 [astro-ph.CO]].
    
\bibitem{consist3} 
P.~ Creminelli, A.~ Perko, L.~ Senatore, M.~ Simonovic and G.~ Trevisan,
  ``The Physical Squeezed Limit: Consistency Relations at Order $q^2$,''
  [arXiv:1307.0503 [astro-ph.CO]].
 

   
   
\end{thebibliography}
\end{document}